\documentclass[reprint,aps,showpacs,prx,superscriptaddress,floatfix,10pt,nofootinbib,twocolumn]{revtex4-2}  	
\usepackage[parfill]{parskip}    		
\usepackage{bibunits}
\defaultbibliographystyle{myapsrev4-2}

\usepackage{amssymb}
\usepackage{amsmath}
\usepackage{graphicx}
\usepackage{bm}
\usepackage{wrapfig}

\usepackage{xcolor}
\usepackage{dcolumn}

\usepackage{array}
\def\Wv{{\mathbf{W}}}
\def\rv{{\mathbf{r}}}

\def\xv{{\mathbf{x}}}

\def\Iv{{\mathbf{I}}}
\def\yv{{\mathbf{y}}}
\def\Av{{\mathbf{A}}}
\def\Bv{{\mathbf{B}}}
\def\Cv{{\mathbf{C}}}
\def\bv{{\mathbf{b}}}
\def\onev{{\mathbf{1}}}
\def\eps{\epsilon}
\def\varphiv{{\bm{\varphi}}}

\def\fv{{\mathbf{f}}}

\def\Fv{{\mathbf{F}}}
\def\epsv{{\bm{\epsilon}}}

\def\Phiv{{\bm{\Phi}}}
\def\Ov{{\mathbf{I}}}

\def\@fnsymbol#1{\ensuremath{\ifcase#1\or \dagger\or \ddagger\or
   \mathsection\or \mathparagraph\or \|\or **\or \dagger\dagger
   \or \ddagger\ddagger \else\@ctrerr\fi}}
    \makeatother

\begin{document}

\title{Excitation-inhibition balance in cortical networks with heterogeneous\\ cluster sizes and its applications}

\author{Abhijit Chakraborty*$^,$}
\affiliation{Department of Physics, University of Houston, Houston TX 77204, USA}
\author{Greg Morrison$^\dagger$}
\affiliation{Department of Physics, University of Houston, Houston TX 77204, USA}
\affiliation{The Center for Theoretical Biological Physics, Rice University,
Houston, TX 77005}

\def\thefootnote{*}\footnotetext{Current affiliation: Institute for Quantum Computing, University of Waterloo, Waterloo, ON, Canada, N2L 3G1}

\def\thefootnote{$\dagger$}\footnotetext{email: gcmorrison@uh.edu }

\date{\today}
\begin{bibunit}
\begin{abstract}
Insight into how information can propagate within cortical networks is essential for a more complete understanding of neural dynamics and computation in complex networks.  Networks with clustered connections have previously been shown to give rise to correlated dynamics in individual clusters. However, this same model applied to a network with highly heterogeneous cluster sizes leads to a clear breakdown of the balanced state. In this article, using a formal definition of the balance matrix, we show why the balance condition breaks and propose a solution to restore balance in heterogeneous networks by reweighing the connection strengths based on community sizes.  We introduce a method of partially balancing a heterogeneous network and show that the degree of spontaneous synchronization within communities can be varied using a single parameter describing the reweighing.   We further show that stimuli can propagate through a hierarchically clustered network, where stimulating one cluster of neurons in a densely connected pair induces correlated firing in the other without propagating to other weakly connected clusters. 
\end{abstract}
\maketitle

\section{\label{sec:1}Introduction}

Information processing and other essential biological functions in the brain are driven by coherent activity in regions of the brain \cite{Salinas:2001jr,Smith:2013fw,Kumar:2010dv,Beggs:2003uv,Atasoy:1ev}, strongly influenced by both external stimuli \cite{Gollisch:2008jv,Ebsch:2018bi,Chettih:2019el} and the connectivity between neurons at short \cite{Smith:2013fw,Rosenbaum:2014ft} and long \cite{Modha:2010ih,Oh:2014bu} distances.  Synchronous \cite{Salinas:2001jr} and asynchronous \cite{Ostojic:2014kd} activity have been shown to play a role in neural coding, and pathological firing dynamics may be associated with diseases such as epilepsy \cite{Jiruska:2013ja,Bower:2012ce} or schizophrenia \cite{schitz}.    Trial-to-trial variability \cite{Bower:2012ce,Rosenbaum:2016hm} in neural firing indicates that this coordination in activity must be a collective property of the network, rather than a deterministic sequential process of individual neurons.  Because different species and different individuals within each species are able to accomplish similar information processing tasks, this biologically essential coordination of neural activity must be highly robust to heterogeneity in network topology \cite{Varshney:2011ju,Oh:2014bu,Chettih:2019el,Swanson:2017dl,Shimono:2015ir,Wu:2011kt,Zhou:2006jc,deReus:2014cz,Giusti:2015kd,Ebsch:2018bi,Klinshov:2014ce,Pyle:2016bvb,LitwinKumar:2012go,Betzel:2017ev} and external stimuli \cite{Ledoux:2011ed,Ebsch:2018bi} in order to accomplish essential tasks.

The topology of neural networks may be highly heterogeneous, and a number of studies have highlighted a variety of indicators in the network science literature as potentially important factors in understanding neural dynamics.  These include the distribution of regional or neuronal degree \cite{Pyle:2016bvb} ($K_c$ and $k_i$ respectively),  the influence of spatial proximity on connections between neurons \cite{Rosenbaum:2014ft,Rosenbaum:2016hm}, and the existence and impact of clusters in cortical networks \cite{Betzel:2017ev,LitwinKumar:2012go}.  Clustered networks refer to those where groups of neurons are more densely connected to each other than they are to neurons outside of the group (this is termed community structure in the network science literature, with both clusters and communities used synonymously in this paper).  A number of experimental studies \cite{Klinshov:2014ce,deReus:2014cz} have shown that cortical networks often adopt complex community structure, with these communities potentially forming a hierarchy \cite{deReus:2014cz,Felleman:lCFH2qwV} through which an external stimulus may pass.   Litwin-Kumar and Doiron have used a computational model to show \cite{LitwinKumar:2012go} that clustered networks can lead to an increase in the activity of all neurons in a single group  either spontaneously or due to direct stimulation.  In addition to community structure in the connections between neurons, communities may themselves form a hierarchy \cite{Betzel:2017ev,Markov:2013it,Morrison:2012jy}, with some communities more densely connected between each other than to the rest of the network.  The ubiquity of complex topologies in cortical networks may play role in the ability to overcome trial-to-trial variability for single neurons by correlating activity of relevant functional groups in computation.

In addition to structural information related to the statistics of connectivity between neurons, networks of neurons should satisfy a condition of balance \cite{LitwinKumar:2012go,Zhou:2006jc} on the level of an individual neuron:  the average excitatory and inhibitory signals from other neurons must be approximately equal for physically realistic neural dynamics.  Unbalanced stimulation of a neuron by its neighbors will lead to hyperactivity or silencing, referred to as mean-driven dynamics, rather than the fluctuation-driven dynamics observed in real brains \cite{Renart:2007eq,Cai:2004em}.  Computational models are generally designed to give balanced dynamics \cite{Rosenbaum:2014ft,Pyle:2016bvb,Deneve:2016hb}, with a balanced state that is stable to perturbations \cite{Pyle:2016bvb} leading to physically relevant neural dynamics.  A classic result regarding balanced neural networks is that the strength of the connection between neurons must \cite{vreeswijk1998chaotic,Ebsch:2018bi} scale as $K^{-1/2}$ (with $K$ the mean degree of the nodes in the Erd\H{o}s-Renyi neural network) in order for fluctuations to persist in the limit of $N\to\infty$.  This scaling is widely used in modeling of neural networks, and indicates the importance of understanding the interplay between network topology and the balance condition.  Not all network topologies are capable of producing balanced firing dynamics, and a more complete understanding of what network topologies and interaction strengths permit balanced firing is essential for realistic modeling of cortical networks.

This paper is organized as follows. In section 2, networks with heterogeneous community sizes are shown to exhibit clearly unbalanced dynamics, representing an unphysical model of neural connectivity.   A strategy to restore balanced firing by reweighing the strength of connection between neurons within and between clusters is discussed in section 3.  In section 4, we apply this procedure to large communities in a network of heterogeneous community size, and show that while balanced firing is indeed observed spontaneous synchronization is completely suppressed. In section 5, we show that a procedure of partial balance (which breaks the balance condition by tuning a single parameter) recovers spontaneous synchronization and permits external stimulation.   Finally, we show that communities-of-communities (where a pair of clusters are more densely connected to each other than other clusters in the network) are capable of exciting each other without significantly perturbing the activity in the rest of the network.  The paper concludes with a discussion of the utility of this approach for better modeling complex cortical networks with heterogeneous topologies.

\section{Activity in heterogeneous network with homogeneous connection strengths\label{modelSec}}

Spontaneous synchronization of neural activity has been observed in Leaky Integrate and Fire (LIF) models for which exciters are divided into clusters \cite{LitwinKumar:2012go}, where exciter neurons are more likely to be connected to other neurons within their cluster than to neurons within other clusters.  This motivates the current study, where we wish to evaluate the effect of significant heterogeneity of cluster sizes using a similar model.   Throughout this paper, we consider a network of 4000 excitatory neurons ($N$) and 1000 inhibitory neurons ($M$) each following a leaky integrate and fire model. The membrane potential of any neuron $j$ is governed by the ordinary differential equation
\begin{equation}
\dot{V}_j = \frac{1}{\tau_j}(\mu_j - V_j) + I_{j,syn} \label{leaky}.
\end{equation}
$V_j$ is non-dimensionalized membrane voltage with threshold voltage $V_{th} = 1.0$ and refractory period following the spike is 5 ms. $I_{j,syn}$ is the synaptic current experienced by neuron $j$. The synaptic current is modeled by the spike trains received by the neuron convoluted with an exponential filter (See Supplementary Information). $\tau$ represents the timescale of firing for a neuron. The timescale ($\tau_j$) for excitatory neurons are 15ms and for inhibitory neurons are 10ms  \cite{LitwinKumar:2012go}. $\mu_j$ is the bias voltage applied to a neuron, drawn from a uniform distribution \cite{LitwinKumar:2012go} between 1.1 and 1.2.

Clustered neural connectivity is defined in terms of a greater density of connection or greater connection probability within a group vs between groups.  In this paper, we assume inhibitory neurons are unclustered, so the connection probability from a neuron in population $k$ to a neuron in population $j$ is denoted by $p_{jk}$ with $p_{ei} = p_{ie} = p_{ii} = 0.5$ (where subscript $e$ denotes the exciter population and the subscript $i$ denotes inhibitor population).  Excitatory neurons are assumed to have a more complex topology of neural connectivity, where clusters of excitatory neurons are more likely to be connected to other neurons in the same cluster than to other neurons.   The parameters $R_p$ (a ratio of probabilities) and $R_J$ (a ratio of connection strengths) define the degree of clustering, with $R_p = \frac{p_{ee}^{in}}{p_{ee}^{out}}$ and $R_J = \frac{J_{ee}^{in}}{J_{ee}^{out}}$.  $R_p = 1$ indicates that there is no density difference and $R_j = 1$ indicates there is no difference in connection strengths within a group vs between groups.  Following  \cite{LitwinKumar:2012go}, the initial values of the parameters are chosen to be $R_p=2.5$, $R_j=1.7$. The connection probability between two excitatory neurons is then calculated such that the degree of connectivity ($K$) of each excitatory neuron is same. This means that on average each exciter is connected to $K=800$ other exciters.   Using these parameters to create a network formed of clusters of homogeneous size, spontaneous synchronized firing and variability inside the communities as established in  \cite{LitwinKumar:2012go} (and reproduced in the  S.I.). The synaptic current in any of the communities shows that the network appears close to the balanced state of equal exciter and inhibitory stimuli. This behavior of the network is robust to mild heterogeneity introduced in the community sizes by choosing a normal distribution centered on the mean.

\begin{figure*}[ht!]
\centering
\includegraphics[width=0.7\linewidth]{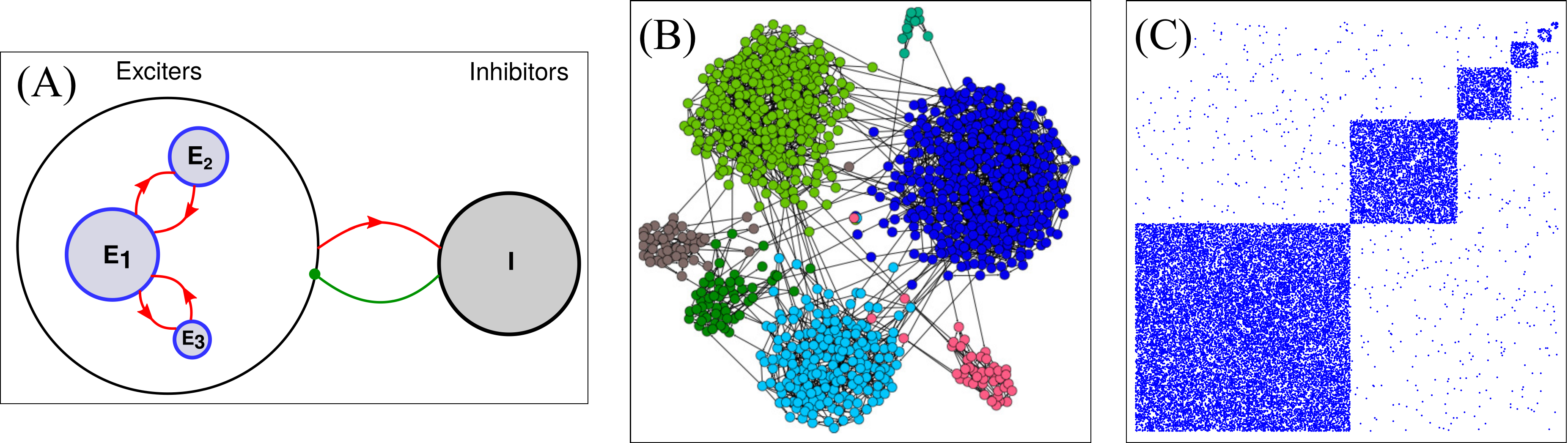}

\caption{Schematics of the heterogeneous networks. (a)  The excitatory population is divided into communities with different sizes, while inhibitors are found in a single community. Arrowheads on connections between groups represent excitatory feedback and rounded heads represent inhibitory feedback. (b) A network schematic of the heterogeneous excitatory network (inhibitors not shown) generated in Gephi \cite{ICWSM09154} for a network of communities whose size satisfy a scale free distribution.  Nodes denote  excitatory neurons and connections the edges between neurons; the weight of each connection is not shown.   (c) shows an adjacency matrix for exponentially distributed community sizes, where blue points denote a connection and white denote no connection.  In either exponential or scale free, the largest community may contain two orders of magnitude greater number of neurons than the smallest community in our simulations.\label{fig:schematics}}

\end{figure*}

\begin{figure*}[ht!]
\includegraphics[width=0.7\linewidth]{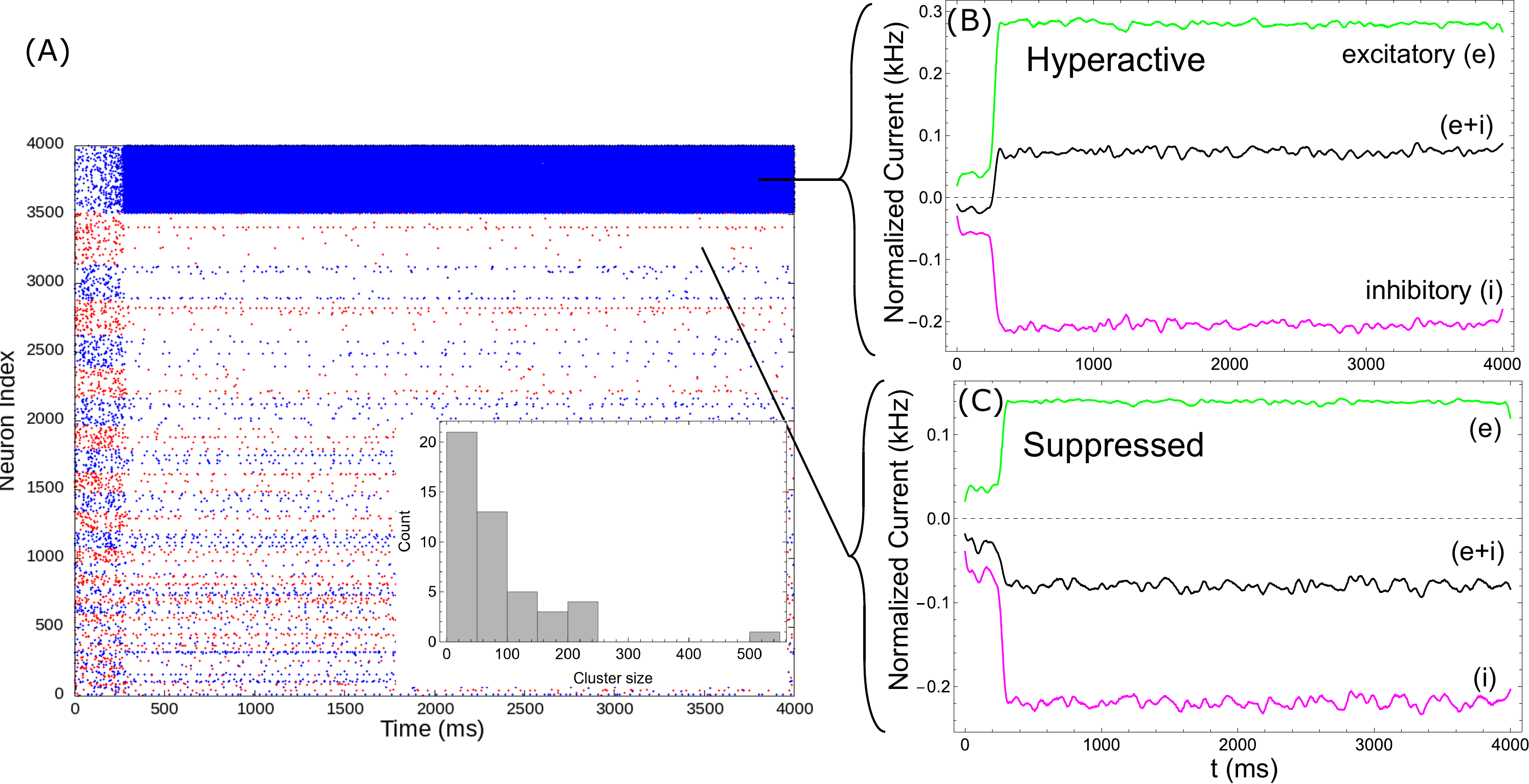}
\caption{Breakdown of balance in heterogeneous communities.  (A) Raster plot of network with exponentially distributed community sizes show a hyperactive largest community and suppressed smaller communities. In the inset the cluster size distribution is shown for this particular realization of the network. (B) Synaptic input currents of representative neuron in the largest community showing high total excitatory input leading to the hyperactivity. (C) Synaptic input current to a neuron in cluster 2 showing high net inhibitory current resulting in suppression of the community.  \label{fig:balancebreaks}}
\end{figure*}

In Fig. \ref{fig:schematics}, schematics of an extremely heterogeneously clustered network are shown. Instead of clusters with equal number of neurons (or with mild heterogeneity with a sharp peak about the mean in the distribution of cluster sizes), heterogeneous networks are constructed of clusters with an scale-free \cite{Kim:2018fb,Li:2010fj} or exponentially \cite{Modha:2010ih} distributed sizes. Using the same parameter values of $R_p$ and $R_j$ as in \cite{LitwinKumar:2012go} with an exponential distribution of community sizes leads to hyper-activity in the largest community and near complete suppression in the smaller communities (Fig. \ref{fig:balancebreaks}).  Other distributions of community sizes that also produces heterogeneous topology (e.g. Gaussian with large variance, power-law) exhibit this same breakdown of balanced state as discussed in the SI.   For some very small communities the firing rate may not be suppressed (and in fact may be hyperactive) due to the sparse connections to inhibitors or to the hyperactive cluster. Thus extreme heterogeneity in community sizes (irrespective of the particular distribution) with the same connection parameters as used in the homogeneous topology results in excess of excitation stimuli or inhibition stimuli in the communities, clearly destroying the balanced synaptic input of the neurons.

The breakdown of balance upon the introduction of heterogeneity calls for a detailed investigation on the balance condition. Even though networks of homogeneous cluster sizes exhibit spontaneous correlated firing, the assumption of homogeneity in cluster sizes is a rather strict requirement. Studies have shown the presence of structural heterogeneity in cortical networks  \cite{lee2016anatomy,sporns2013structure,Betzel:2017ev} and the failure of the same method to address this more general case raises an important question:  is there a way to avoid this failure to maintain a balanced dynamics in a heterogeneous network?     

%

\section{Strategy to restore balance}
\label{BalanceSec}
The breakdown of balance for a network with heterogeneous community sizes can be understood using the formalism of \cite{Rosenbaum:2014ft}. Using a mean field approach, the synaptic current in each population can be written as
\begin{equation}
\mathbf{I} = \mathbf{W} \cdot \mathbf{r} + \mathbf{F}, \label{current_meanfield}
\end{equation} 
where $\mathbf{I}$ is the mean synaptic input current for each population, $\mathbf{r}$ is the mean firing rate of the clusters, and $\mathbf{F}$ is the supra-threshold bias current. $\mathbf{W}$ is the mean-field {balance matrix} whose elements are given by
\begin{equation}
\mathbf{W}_{jk} = N_k \langle J_{jk} \rangle\qquad \langle J_{jk} \rangle = p_{jk} J_{jk}, \label{mat_elements}
\end{equation} 
with $j,k\in\{1,2,\dots C\}$ for a total of $C$ clusters in the excitatory population. We assume the set is ordered from largest to smallest (so that excitatory community 1 is larger than 2 and so on), and define community $C+1$ as composed of the inhibitory neurons (having no additional community structure).  $\mathbf{W}_{jk}$ represents the average strength of the connection from neurons in group $k$ to neurons in group $j$. The balanced state in the network can be achieved in the mean field limit if the synaptic current is very small and hence from eq \ref{current_meanfield}, $\mathbf{W} \cdot \mathbf{r} + \mathbf{F} \approx 0$. For firing rates $\mathbf{r}$ to be finite, the balance matrix $\mathbf{W}$ has to be non-singular. However, for the balanced state to be a stable one we need to consider the dynamical mean field equation \cite{vreeswijk1998chaotic, dayan2001theoretical, ermentrout2010mathematical}  
\begin{equation}
\tau \mathbf{\dot{r}} = - \mathbf{r} + f(\mathbf{W} \cdot \mathbf{r} + \mathbf{F}) \label{dynamical firing rate eqn}
\end{equation}
For LIF models we can assume that the function $f(\cdot)$ is a threshold linear function, which is usually taken to be a sigmoid function (See Supplementary Information). With this approximation, stable balance can be obtained if all eigenvalues of the matrix $\mathbf{W}$ has a negative real part  \cite{Rosenbaum:2014ft,Pyle:2016bvb}. 
The balance matrix for a network with homogeneous community size is
\begin{equation}
\mathbf{W}_{hom} = \begin{pmatrix}
a N_0 & b N_0 & bN_0  & \cdots & bN_0  & -cM\\
bN_0  & aN_0  & bN_0  & \cdots & bN_0  & -cM\\
\vdots & \vdots & \vdots & \cdots & \vdots & \vdots\\
bN_0  & bN_0  & bN_0  & \cdots & aN_0  & -cM\\
dN_0  & dN_0  & dN_0  & \dots & dN_0  & -eM
\end{pmatrix}, \label{W_homo}
\end{equation}  
where $N_0=N/C$ is the number of neurons in each homogeneous community, $M$ is the number of inhibitors, and where the mean field interaction strengths per neuron are $a = J_{ee}^{in}p_{ee}^{in}$, $b = J_{ee}^{out}p_{ee}^{out}$, $c = |J_{ei}|p_{ei}$, $d = J_{ie}p_{ie}$, and $e = |J_{ii}|p_{ii}$.  Note that $a-e$ should satisfy the condition of scaling as $\sim K^{-1/2}$ if $N$ were to be varied for the balanced condition to be preserved \cite{vreeswijk1998chaotic,Ebsch:2018bi}, but in this paper we focus solely on a fixed value of $N$.  It is readily verified that this balance matrix has $(C-1)$ equal eigenvalues with the stability condition given by
\begin{subequations}
\begin{eqnarray}
&a-b<0 \qquad({\mbox{with degeneracy }} C-1), \label{eigcondition}\\
&a + (C-1)b > e,\\
&\frac{a+(C-1)b}{Cd}e < c <\frac{(a + (C-1)b + e)^2}{4Cd}.
\end{eqnarray}
\end{subequations}
The condition in eq \ref{eigcondition} implies that, a perfectly balanced network requires $\frac{J_{ee}^{in}}{J_{ee}^{out}}<\frac{p_{ee}^{out}}{p_{ee}^{in}}$. Deviation from this condition gives rise to more complex firing dynamics within each group, including chaotic or unstable state. In the homogeneous network with parameters described in the previous section ($R_p = 2.5$, $R_j = 1.9$), the balance condition is not satisfied. Both the connection strength and the connection probability are greater inside the communities than outside. The spontaneous correlated firing in the homogeneous network is the result of the failure to meet the condition in eq \ref{eigcondition}. But for the homogeneous communities or mildly heterogeneous networks (community sizes sharply peaked around mean value) the effect of the imbalance does not lead to overwhelming overstimulation or suppression of any community. In the case of largely heterogeneous network, the violation of the balance criteria gives rise to a complete breakdown of balance.

In a network with heterogeneous community sizes the balance matrix takes the form
\begin{equation}
\mathbf{W}_{het} = \begin{pmatrix}
aN_1 & bN_2 & bN_3 & \cdots & bN_C & -cM\\
bN_1 & aN_2 & bN_3 & \cdots & bN_C & -cM\\
\vdots & \vdots & \vdots & \cdots & \vdots & \vdots\\
bN_1 & bN_2 & bN_3 & \cdots & bN_C & -cM\\
dN_1 & dN_2 & dN_3 & \cdots & dN_C & -eM\\
\end{pmatrix}\label{W_hetero}.
\end{equation}
where $a,b,c,d,$ and $e$ are the same in the homogeneous case.  This weight matrix produces the non-physical firing rates observed in Fig. \ref{fig:balancebreaks}, which begs the question:  given a heterogeneous connection probability $p_{ij}$, for what values of the connection strengths is balanced firing possible?  In this paper, we refer to adjusting the connection strengths from some initial value $J_{ij}$ to a new value $J_{ij}'$ as `restoring balance.'

To restore balance in the network, we adjust the connection strength to make the eigenvalues of $\mathbf{W}_{het}$ negative, with the goal of producing a mean field weight matrix that will produce physically meaningful firing rates given a specific network topology.  Although it should be possible to alter the pre- or post-synaptic inhibitory weights in such a way that the criteria of all negative eigenvalues is met, it becomes dauntingly difficult to determine a tractable method to do so with large number of excitatory communities. Analytical determination of eigenvalues beforehand is also difficult for large $C$ (as discussed in the SI). However, there is a trivially simple way to redefine the weights which satisfies the balance condition:  recast eq. \ref{W_hetero} in the symmetric form of eq. \ref{W_homo} by modifying the weights to a new value $J_{jk}' \propto \frac{J_{jk}}{N_k}$ for all excitatory clusters (those with $k\le C$).  That is, the interaction strength originating from excitatory neuron is reduced proportional to the size of the excitatory community of which it is a member. The advantage of this simplistic rescaling is that the $(C-1)$ degenerate eigenvalues are known analytically from the matrix in eq \ref{W_homo} and the balance criteria can be satisfied simply by ensuring $b>a$. Fulfilling the criterion $b>a$ implies choosing $R_p$ and $R_j$ in such a way that the original connection strengths satisfy $\frac{J_{ee}^{in}}{J_{ee}^{out}}<\frac{p_{ee}^{out}}{p_{ee}^{in}}$.  Dividing the connection strengths by the community sizes reduces the overall strength going in and out of each community in a manner satisfying the balance condition.   This procedure produces a balanced matrix $(\Wv')_{jk}=N_kJ_{jk}'p_{jk}$ for which (a) all connection probabilities are the same as in \ref{W_hetero} and (b) all of the eigenvalues of $\Wv'$ are all negative.  

The constant of proportionality in $J_{jk}'\propto \frac{J_{jk}}{N_k}$ remains to be determined using the procedure outlined above.  After reducing the strength of the connections within the communities, the total pre-synaptic strength of the network   has been significantly reduced by this procedure.  Defining $s_k=\sum_{j\ne k} W_{jk}$ and $s'_k=\sum_{j\ne k}W_{jk}'=N_k^{-1}s_k$, we see that the pre-synaptic weight of each community is reduced by its size $N_k$, and thus defining $S(C)=\sum_{k=1}^{C}s_j$ and $S'(C)=\sum_{k=1}^{C}s'_j$, it is readily seen that $S'(C)/S(C) \ll 1$ for large networks.  Simply normalizing each community by its size will thus significantly reduce activity in comparison to the homogeneous network, and we expect that we must choose $J_{jk}'=\varphi(C) J_{jk}/N_k$ with $\varphi(C)=S(C)/S'(C)$ to produce a firing rate consistent with the homogeneous network.  However, this approach produces unrealistic neural firing patterns as well for a different reason:  communities of very small size are given enormous interaction strengths with other communities (since $N_k\ll \varphi(C)$ for small communities $k$), and the firing within the network becomes synchronized with these small clusters.  In order to overcome this problem, we chose to rescale the weights belonging only to sufficiently large communities (the method of selecting the cutoff is described in the Supplementary Information.  In the exponentially distributed network sizes shown in the figures below, this cutoff was chosen for $C^*=25$, with 
\begin{eqnarray}
\Wv'_{jk}=N_kJ_{jk}'p_{jk}\qquad\qquad J_{jk}'=\left\{\begin{array}{cc}\frac{\varphi(C^*)}{N_k}J_{jk} & k\le C^* \\J_{jk} & k>C^*\end{array}\right.\label{actualBalance}
\end{eqnarray}
Note that the presynaptic weights from inhibitory neurons, which lack any community structure, are left unaltered using this procedure (since $k=C+1$ for the inhibitory cluster).  

%
%
\begin{figure*}[ht!]
\includegraphics[width=0.7\linewidth]{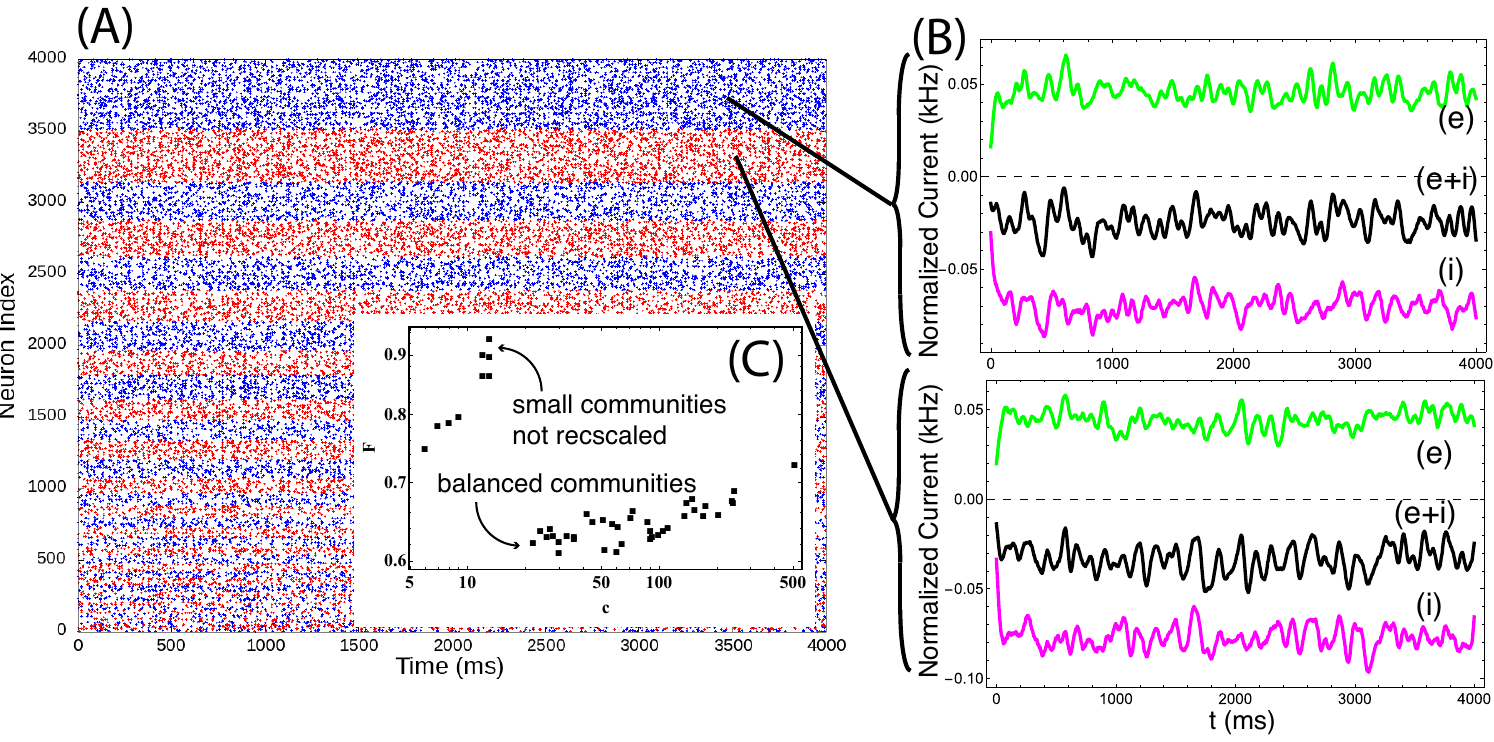}
\caption{(A) Raster plot of the rebalanced network following procedure described in section \ref{Result1}. Hyperactivity and hypersuppression are not evident after rebalancing as they were in Fig. \ref{fig:balancebreaks}(A). (B) shows the current for a representative in the largest (top) and second largest (bottom) communities, which both show virtually indistinguishable current dynamics.  (C) shows the Fano factor for each community as a function of community size, with the larger  communities having been rebalanced and the smaller communities left unbalanced (described in the SI). \label{fig:nosync}}
\end{figure*}

\section{Balanced heterogeneously clustered networks \label{Result1}}

Having rescaled the presynaptic strength of each excitatory neuron proportional to the community size (as described in the section above)), the firing dynamics (seen in Fig. \ref{fig:nosync}(a)) shows that the hyperactivity previously seen in \ref{fig:balancebreaks} is no longer present.  Rebalancing the weights has also recovered a balanced state for the neurons in the large community (shown by the traces in Fig. \ref{fig:nosync}(B)), with the excitatory and inhibitory signals near zero for all neurons in the larger communities.  Fig. \ref{fig:nosync}(C) shows the Fano factors $F_i$ of the neurons within community $i$ as a function of community size, with $F_i=N_i^{-1}\sum_{n\in C_i} v_n/r_n$, where the variance $v_n$ of any neuron $n$ in community $i$ is normalized by that neurons firing rate $r_n$ (the rate and variance were estimated over 100ms intervals).  The Fano factor is precisely 1 for a Poisson distribution and is precisely 0 for a constant, so communities with $F_i>1$ can be considered as having high variability \cite{LitwinKumar:2012go}.  The variability of the clusters has a weak dependence on the community size (ranging between $F_i\approx 0.6-0.7$, but clearly shows the Fano factors are below $F_i=1$ for all communities.  A sharp difference is found for the Fano factors of unbalanced communities (those with $C_i<15$, having $F_i\approx 0.75-0.95$), but the variance still remains lower than what would be expected for a Poisson distribution.

\begin{figure}[h]
\centering
\includegraphics[width=1.0\linewidth]{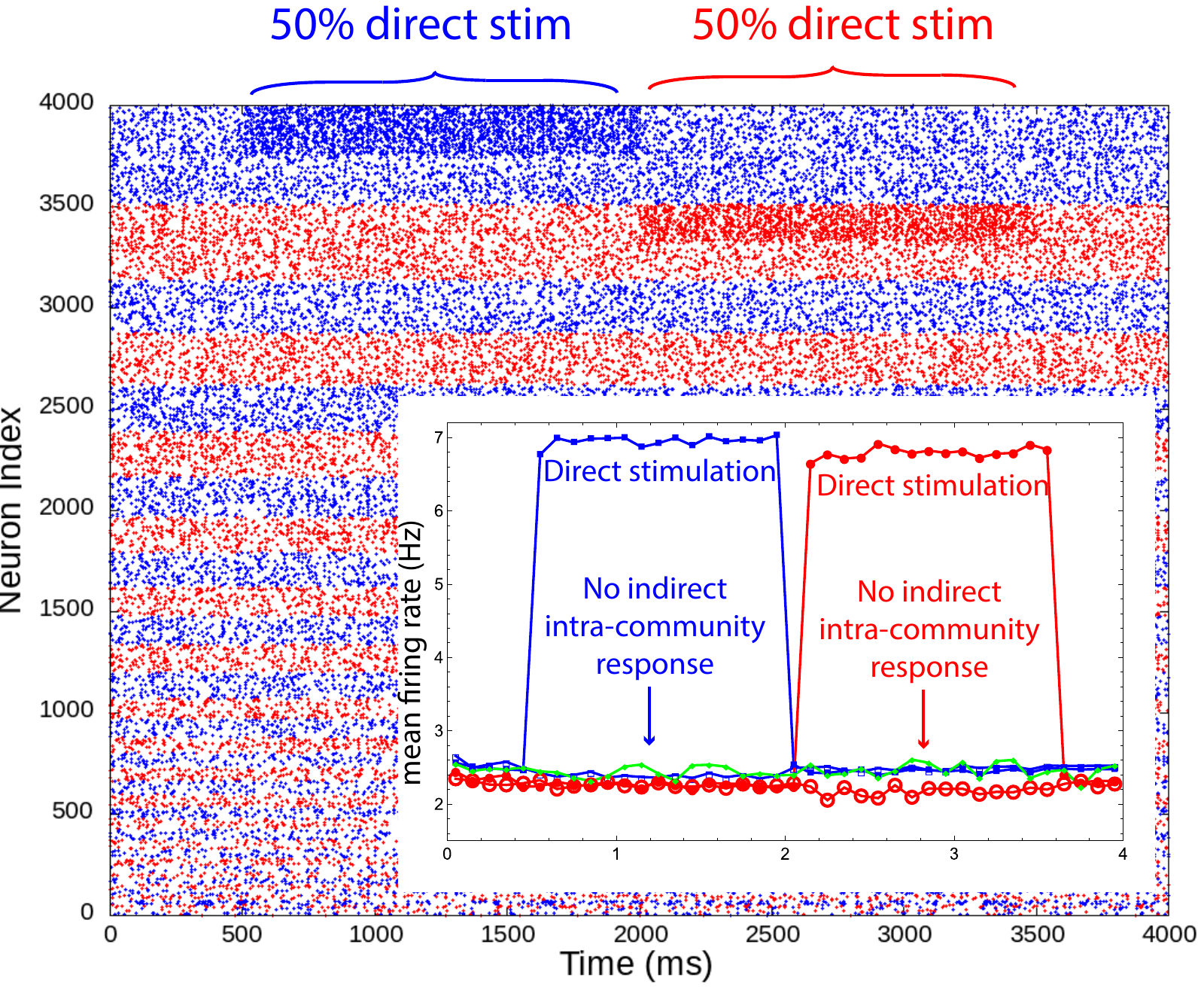}
\caption{(A) Raster plot of direct stimulation of 50\% of the neurons in the largest community from $t=$500 to 2000 ms and in the second-largest community for $t=$ 2500 to 3500 ms produce increased firing in only half of the clusters, with no indirect stimulation of any other neuron in the community. (B) Firing rates confirming the activities in cluster 1 and 2 during the simulation, with blue solid squares denoting the firing rate of directly stimulated neurons in cluster 1, and open squares denote the firing rate of other half of the community, and solid and open red circles representing the same thing but in cluster 2. The green solid diamonds represent firing rate of a randomly chosen balanced community which does not receive any stimulation.}
\label{fig:stimperfect}
\end{figure}

To determine the ability of the balanced network to propagate excitement within a community, we perform a simulation with the synaptic current increased by a constant bias for 50\% of the neurons in the two largest clusters with the expectation that the remaining unstimulated neurons in the community would experience correlated firing due to the community structure.  In Fig. \ref{fig:nosync}, we see the rebalancing procedure has (perhaps surprisingly) completely suppressed the spontaneous correlated firing inside the communities that was previously observed for homogeneous communities \cite{LitwinKumar:2012go}.  This is due to the fact that normalizing the connection strengths by community size has effectively removed any meaningful community structure in the network:  while the connection probability is higher within than between clusters, $\Wv_{ii}'<\Wv_{ij}'$.  That is, the effective strength of interaction within a community is weaker than the effective strength between a community using this reweighing procedure.  In a truly balanced state each neuron within a community receives the same amount of excitatory and inhibitory stimulation, whereas the correlated activity is driven by an excess of local excitatory stimulation from within a community.   The failure of the balanced state to excite a community can be clearly seen by applying a direct external stimulus to a cluster, shown in  Fig. \ref{fig:stimperfect}.  Direct stimulation of 50\% of the neurons within the largest or second-largest clusters do cause a significant increase in their activity, but do not excite other neurons in the same cluster.  For a perfectly balanced network, neural computation within a clustered community appears impossible as stimuli cannot effectively propagate through a cluster.


The failure of direct stimulation of a subset of community to excite other nodes in that same group clearly indicates that our procedure for enforcing balance not only removes the interesting features of spontaneous synchronization, but also prevents external stimuli from propagating within a community. Rebalancing the network as prescribed by eq \ref{actualBalance} thus completely removes the possibility of neural coding in clustered network. One immediately may wonder whether this is solely an artifact of the rescaling procedure described in eq \ref{actualBalance} and if some other procedure will permit spontaneous synchronization. In the SI we show that for some values of $a,b,c,d, \mbox{ and } e$, it is impossible to balance the matrix without reducing the self interaction weight $a$ (i.e. it is impossible to have all negative eigenvalues choosing inhibitor connection strength freely but $a,b$ fixed). We therefore expect that even if it were possible to re-scale the elements of the balance matrix to produce negative eigenvalues and spontaneous correlated firing, such a procedure would be possible only for a limited parameter space.

  \section{Partially balanced cortical networks \label{Partial Balance}}

\subsection{Partial balance\label{partialBalanceSec}}

 Enforcing balance through the procedure described in the previous section removes the possibility of spontaneous synchronization in the communities (as seen in Fig. \ref{fig:stimperfect}). On the other hand, the failure to enforce balance for heterogeneous communities produces physically unrealistic dynamics (as seen in Fig. \ref{fig:balancebreaks}).  In order to model the dynamics of heterogeneously clustered cortical networks that produce physically meaningful firing rates, we must create a ``middle ground'':   the connection strengths should be scaled such that the matrix prevents hyperactivity but far enough from balance to permit synchronization and propagation of stimulation.  We can accomplish this by increasing intra-community strengths, $J_{ee}^{in}$, relative to the inter-community strengths $J_{ee}^{out}$, moving beyond the balance condition of $J_{ee}^{in}/J_{ee}^{out}<p_{ee}^{out}/p_{ee}^{in}$.  Note that this is equivalent to modifying the weight matrix further, with the addition of a diagonal matrix $\Wv_p$ perturbing the interaction strengths within each cluster of excitatory neurons.

 

  \begin{figure*}[!ht]
  \centering
  \includegraphics[width=0.7\linewidth]{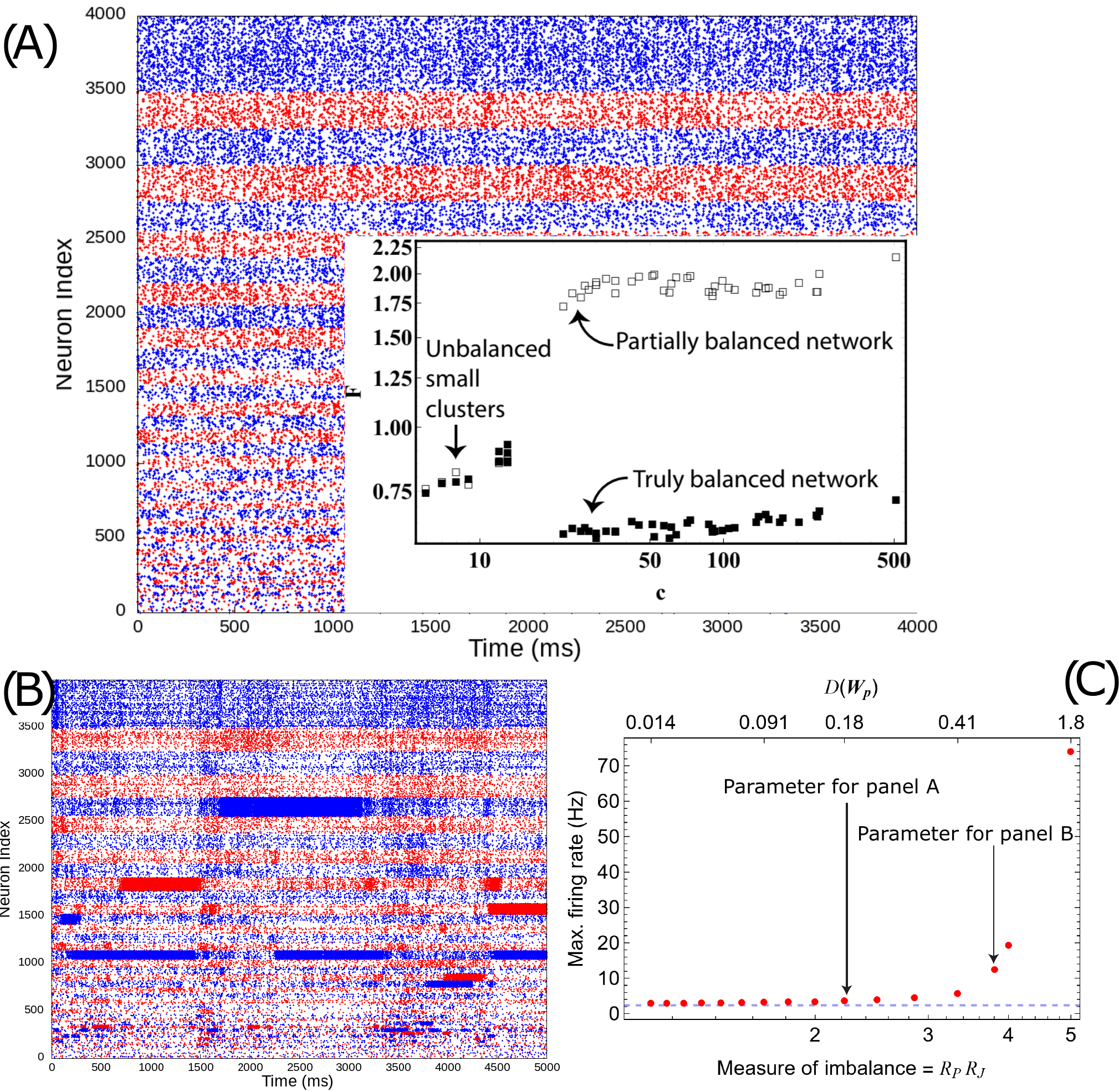}
  \caption{(a) Raster plot for partially balanced, unstimulated network with $D(\Wv_p)=0.2$.  (B) shows the raster plot when $D(\Wv_p)=0.55$, and clearly shows spontaneous synchronization in multiple communities.   Note that spontaneous activity is distributed across both large and small communities.  (C) shows the firing rate as a function of $R_PR_J$ and $D(\Wv_p)$ ($x$-axis on the top) both of which can act as a measure of imbalance in the circuit. The $y$-axis is the average maximum fire rate of a network (considering only communities on which the balancing procedure has been applied). The average has been taken over 50 runs for each parameter value $R_PR_J$ which is enough to get error bars small enough to be indistinguishable from the data points. The firing rate rapidly diverges when $J_{ee}^{in}/J_{ee}^{out}\gg p_{ee}^{out}/p_{ee}^{in}$. The dotted line represents the average firing rate of a perfectly balanced network. The inset of (A) shows the Fano factors for the communities as a function of their size for $D(\Wv_p)=0.54$ (empty symbols) and for the perfectly balanced network (filled symbols), identical to Fig. \ref{fig:nosync}(C).}
  \label{fig:partial_balance_unstim}
  \end{figure*}

The addition of the diagonal matrix to the balance matrix is treated as perturbation and causes a change in the firing rate which will permit correlated firing.  To determine the effect this perturbation, we modify the mean field equation in Eq. \ref{dynamical firing rate eqn} for the firing rate of cluster $i$, $\rv_i$, given by  $\tau \dot\rv=-\rv+f(\Wv'\cdot\rv+\Fv) $.    
After perturbation of the intra-cluster connection strengths, the firing rate equation becomes $\tau \dot\rv^\prime = -\rv^\prime+f((\Wv'+\Wv_p)\cdot\rv^\prime+\Fv)$ where $\rv^\prime$ is the perturbed firing rate.  Rewriting $\rv^\prime = \rv + \epsv$, where $\epsv$ is the change in firing rate due to the introduction of the imbalance, to first order we find
  \begin{equation}
  \tau \dot{\epsv} = -\epsv + [\Wv'\cdot\epsv + \Wv_p \cdot (\rv+\epsv)]f^\prime(\Wv'\cdot\rv+\Fv) \label{eq:epsfirstorder}.
  \end{equation}
  In steady state $\dot{\epsv} = 0$ and $\Wv'\cdot\rv+\Fv$ is the mean synaptic current received by each community in the perfectly balanced network. Defining $\varphiv_{ij} = \delta_{ij}f'\left( (\Wv'\cdot\rv)_i+F_i\right)$, eq. \ref{eq:epsfirstorder} at steady state can be written as,
  \begin{equation}
  \epsv = [\Iv - (\Wv'+\Wv_p)\varphiv]^{-1}\Wv_p\varphiv\cdot \rv.\label{eq:eps}
  \end{equation}
  with $\rv_i$ the mean synaptic rate for the $i^{th}$ cluster for the perfectly balanced network.  
    
  Eq. \ref{eq:eps} gives an upper bound on the permissible change in the balance matrix with the quantity $\eps = |\epsv|$ giving the magnitude of the change in the firing rate due to the perturbation. If $\eps$ is high with respect to the firing rate of the perfectly balanced network, then the perturbation will drive the dynamics far from the balanced state. The value of $ D(\Wv_p)\equiv {\eps}/{|\rv|} $ quantifies the degree to which the matrix has been driven away from the balanced state, with $D(\Wv_p) = 0$ being perfectly balanced and for sufficiently large $D(\Wv_p)$ nonphysical dynamics may occur.   While we focus on a diagonal perturbation in this paper, we expect off-diagonal perturbations can be incorporated in a similar fashion so long as $D(\Wv_p)$ is sufficiently small.  In the results below, we simply increase the value of $J_{ee}^{in}$ in our simulation (moving beyond the balance constraint of $J_{ee}^{in}/J_{ee}^{out}<p_{ee}^{out}/p_{ee}^{in}$).

\begin{figure}[!ht]
  \centering
  \includegraphics[width=0.8\linewidth]{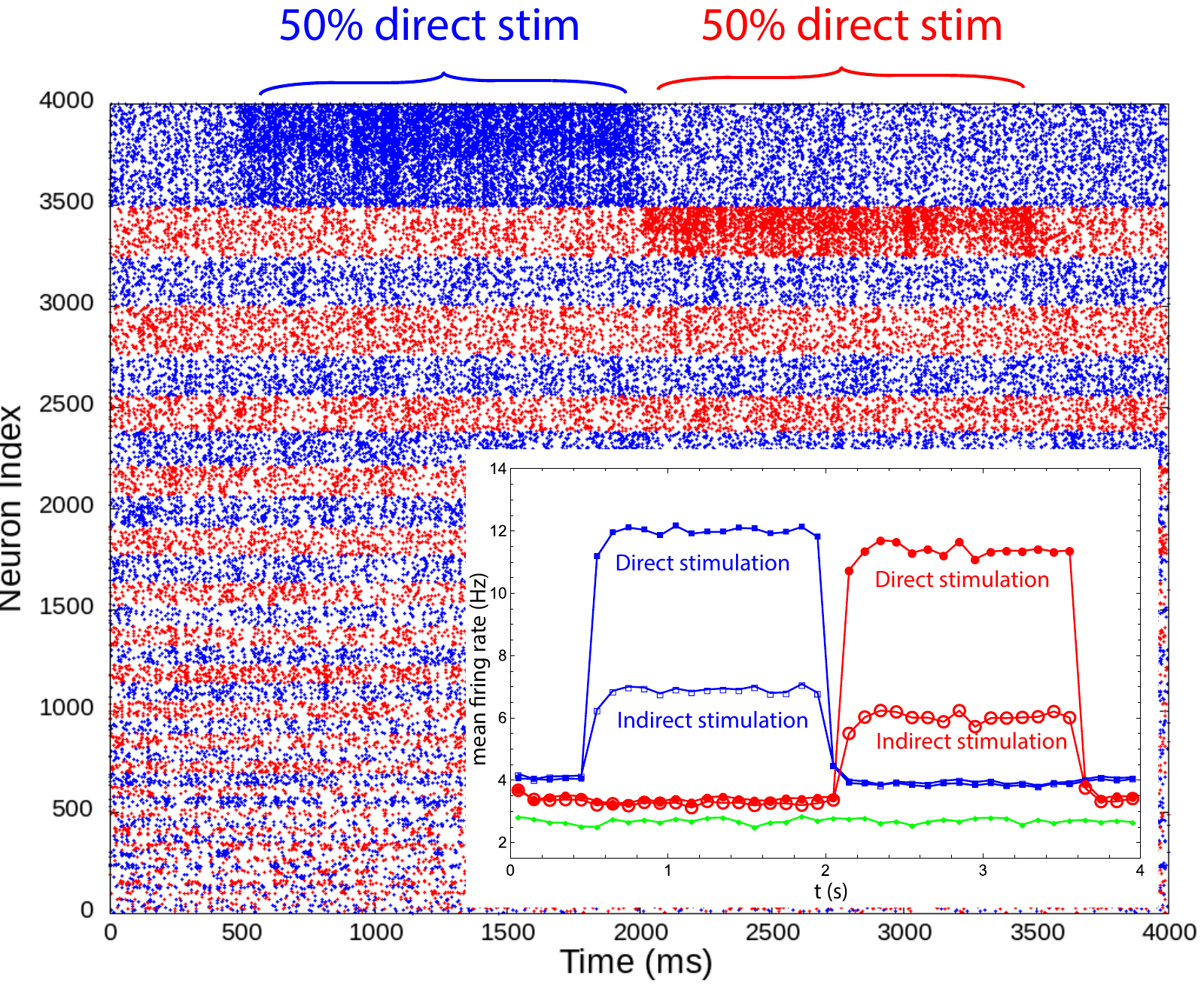}
  \caption{(a) Raster plot for partially balanced network. Stimulation is provided to 50\% neurons of cluster 1 from 500 ms to 2500 ms and to 50\% of cluster 2 from 2500-3600 ms. (b)Firing rate vs. time plot for the partially balanced network. Solid and open blue squares represent directly stimulated fraction of neurons and the unstimulated fraction of neurons respectively in the largest cluster. Red solid and open circles represents the same for the second largest cluster. The green circles represent the firing rate of a community which is never stimulated during the simulation run-time. Each data point represents the firing rate calculated over a 100ms window.}
  \label{fig:partial_balance_stim}
  \end{figure}

Fig. \ref{fig:partial_balance_unstim}(A) shows the firing dynamics an out-of-balance network with $J_{ii}^{in}/J_{ee}^{out}\approx0.6>p_{ee}^{out}/p_{ee}^{in}=0.25.$, violating the balanced condition on the mean field level.  This corresponds to a small perturbation strength $D(\Wv_p) \approx 0.2 \ll 1$ , and we again see that the re-balanced network shows neither hyperactivity (as was seen in Fig. \ref{fig:balancebreaks}) nor spontaneous synchronization (as in \cite{LitwinKumar:2012go}).   By increasing $D(\Wv_p)$ the network can be driven to exhibit correlated dynamics, as pictured in Fig. \ref{fig:partial_balance_unstim}(B).  With $D(\Wv_p)=0.54$ ($J_{ee}^{in}/J_{ee}^{out}\approx 0.95$), spontaneous synchronization is clearly exhibited for heterogeneous communities sizes for both small and large clusters.  The network with homogeneous cluster sizes in \cite{LitwinKumar:2012go} that produces spontaneous correlated firing (with $R_p =2.5, R_j = 1.9$) has $D(\Wv_p) \approx 0.53$, so the same degree of perturbation as characterized by the parameter $D(\Wv_p)$ can produce high variability in a network with heterogeneous community sizes. This is in contrast to the other cluster imbalance parameter $R_PR_J$. In the later case, as we have already seen the same value of $R_PR_J$ can lead to hyperactivity in clusters with big community sizes whereas it leads to spontaneous synchronization in clusters with small/homogeneous communities. So, $D(\Wv_p)$ is a better measure of imbalance in a clustered network which does not depend on the size of the clusters. The Fano factors for a partially balanced system confirm the high variability of partially balanced networks (shown in the inset of Fig. \ref{fig:partial_balance_unstim}(A)).  However, a sufficiently large perturbation rapidly increases the firing rate within the network, eventually leading to the hyperactive behavior seen in Fig. \ref{fig:balancebreaks}, shown in Fig. \ref{fig:partial_balance_unstim}(C).

  \begin{figure}
  \centering
  \includegraphics[width=1.0\linewidth]{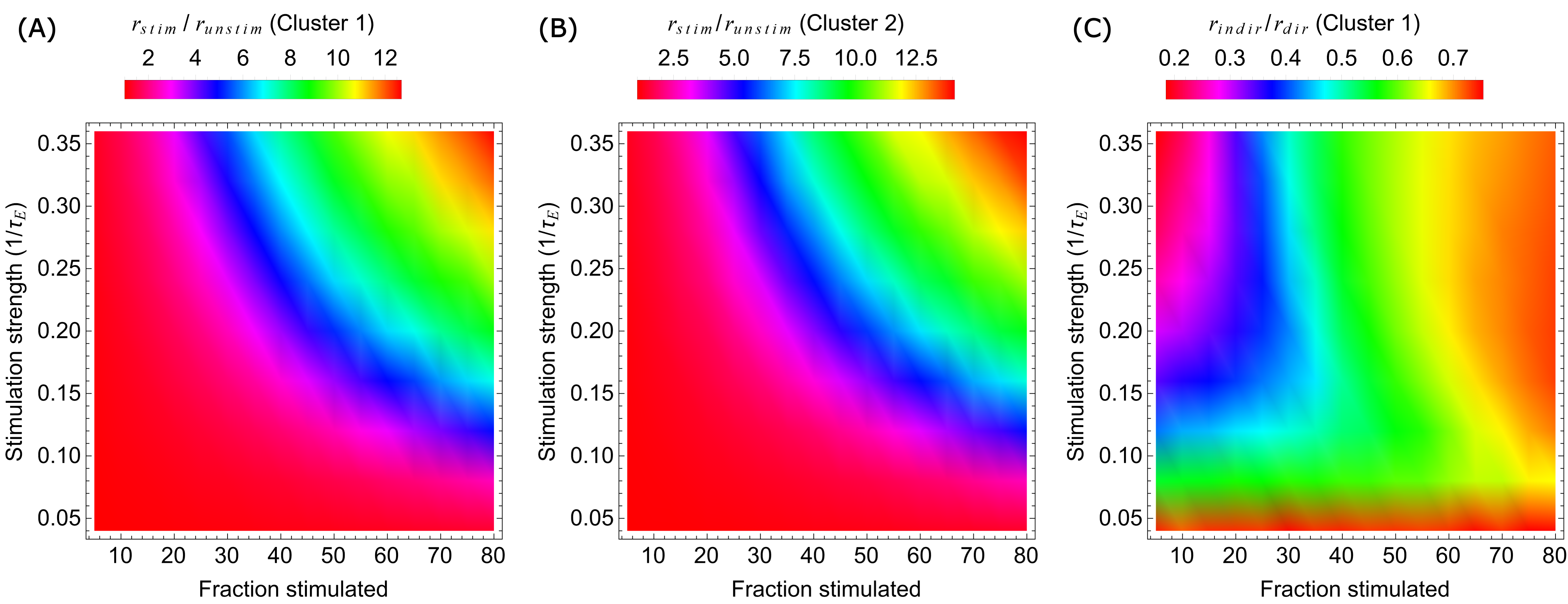}
  \caption{(a) Plot of ratio of the firing rate of the fraction in community receiving indirect stimulation while being stimulated vs while unstimulated. (Stimulation applied to cluster 1). (b) Same plot but now the stimulation is applied to the second largest cluster. The nature of the plot remains the same as (1), showing that the response is not cluster specific. (c) Ratio of the firing rates of fraction of cluster 1 that is directly stimulated vs that of the fraction which receives indirect stimulation.}
  \label{fig:densityplots}
  \end{figure}

\subsection{Stimulation of partially balanced networks}
 When a subset of network is directly stimulated a partially balanced network can show clustering activity, shown in Fig \ref{fig:partial_balance_stim}.  A stimulus provided to a fraction of a single cluster will be propagated throughout the whole cluster,  unlike the firing rates observed for a perfectly balanced network as in Fig. \ref{fig:stimperfect}. The propagation of the stimulus lead to increased activity throughout the cluster (stimulated and unstimulated alike) only for the duration of the stimulation.  Indirectly stimulated neurons show less activity than those that are directly stimulated in the same community, and there is no apparent reduction in the activity of other neurons in the rest of the network (shown in the inset of Fig. \ref{fig:partial_balance_stim}).  The response of the directly stimulated neurons and the secondary response of neurons that were not stimulated are comparable for communities of different size (cluster 1 and cluster 2 respectively). After the stimulation ends, activity in the cluster returns almost immediately to random uncorrelated firing for $D(\Wv_p)=0.2$.

To better understand a partially balanced network's ability to propagate stimuli within a community we vary the stimulated fraction $\rho_{stim}$ and the stimulus strength (in units of $\tau_e^{-1}$, the timescale for excitatory neurons in eq. \ref{leaky}) for fixed $D(\Wv_p) = 0.2$.  One measure of the propagation of stimulation within the community is the ratio of firing rates of the unstimulated fraction during the time period of stimulation and in the absence of stimulation, ${r_{stim}}/{r_{unstim}}$.  This quantity, shown in Fig. \ref{fig:densityplots}(A-B) for the two largest communities, shows that a weak direct stimulation (below $0.1\tau_e^{-1}$) or small fraction of stimulated neurons (below 20\%) are incapable of significantly exciting activity to unstimulated neurons in the same cluster.  Increasing either parameter leads to a greater propensity for indirect stimulation within the cluster, with an increase in the firing rate of over an order of magnitude for indirectly stimulated nodes for high fraction and high strength.

An alternate measure of the propagation of activity is the ratio of the firing rates of the unstimulated fraction and directly stimulated fraction, ${r_{indir}}/{r_{dir}}$, shown in Fig. \ref{fig:densityplots}(C).  The firing rate of indirectly stimulated nodes never exceeds $r_{indir}\approx 0.75r_{dir}$  in our simulations, and this maximal propagation of the stimulation occurs only when $\gtrsim 70\%$ of the neurons with the community are directly stimulated.  For more modest fractions of directly stimulated neurons, the indirect response is $\approx 50-60\%$.

%

\subsection{Stimulation in hierarchically clustered networks}

\begin{figure*}[!ht]
  \centering
  \includegraphics[width=\linewidth]{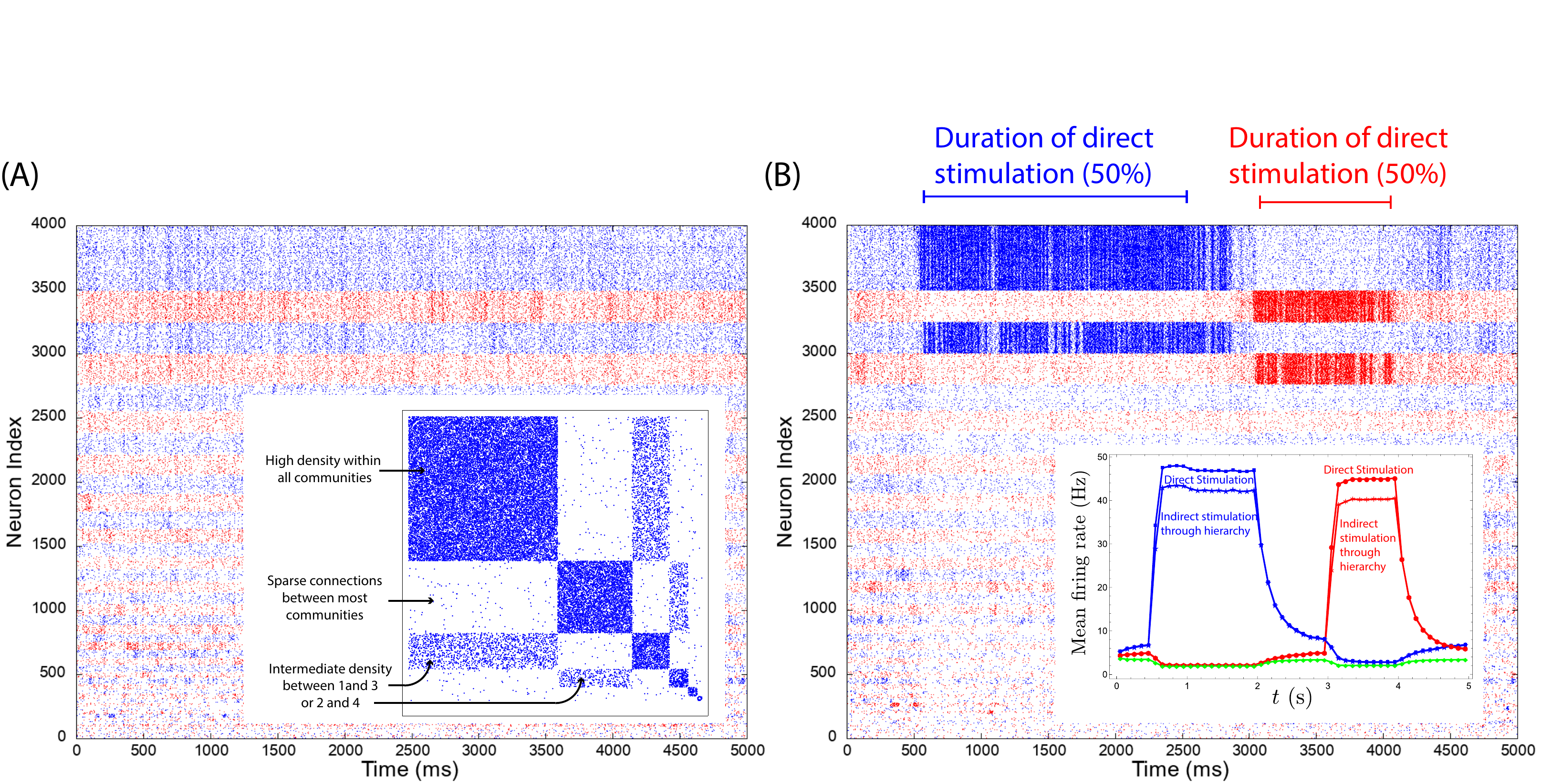}
  \caption{(A)  Raster plot for a network with hierarchy structure. Cluster 1 and cluster 3 belong to one hierarchy, and cluster 2 and cluster 4 belong to another hierarchy.  A schematic diagram of the network topology is shown in the inset of (A), with density differences and community sizes increased for visual clarity.  All intra-community interaction strengths are $\sim 8\%$ weaker than in Fig. \ref{fig:partial_balance_unstim}.  (B) From 500-2500 ms 50\% of cluster 1 is stimulated and from 3000-4000 ms 50\% of cluster 2 is stimulated; in both cases the other member of the hierarchy is excited.  The excitement persists after the direct stimulation is ceased (at $t=2.5$s and 4s).  }
  \label{fig:hierarchy}
  \end{figure*}

Thus far, we have focused on a network exhibiting community structure with heterogeneous sizes.  Heterogeneity in the connections {\em{between}} communities may also occur in real neural networks.   In many contexts, passing signals between communities may also be essential \cite{Felleman:lCFH2qwV,Swanson:2017dl,Zhou:2006jc,Betzel:2017ev}, such as the information processing performed by the visual cortex \cite{Felleman:lCFH2qwV,Kumar:2010dv,Markov:2013it,Bastos:2012bd}, and it is important to assess the ability of a partially balanced network to propagate signals through a hierarchy of communities.  In a hierarchical network, a collection of densely connected clusters of nodes are also more densely connected with each other than to other nodes in the network \cite{Betzel:2017ev,Morrison:2012jy} (forming a community-of-communities structure depicted in the inset of Fig. \ref{fig:hierarchy}(A)).   In a network with hierarchical community structure, there is the greatest connection probability within a community, an intermediate probability of connection between communities in the same hierarchy, and the smallest connection probability between communities in different hierarchies.  In such a network topology, one might naturally expect that excitement is more readily passed within an excitatory cluster, but stimulation of one cluster may excite other clusters in the same hierarchy. 

As a minimal model for this, we construct a network with communities of exponential size as described in Sec \ref{modelSec} and diagrammed in the inset of Fig. \ref{fig:hierarchy}(A):  each neuron is connected between others in its community with probability $p_{ee}^{in}$ and to other excitatory neurons outside of their hierarchy with probability $p_{ee}^{out}=0.4p_{ee}^{in}$.  For this simple model of distinct hierarchies, we connect neurons in the largest community ($C_1$) to the third-largest ($C_3$) with probability $p_{ee}^{mix}=0.75p_{ee}^{in}>p_{ee}^{out}$ (and similarly for the second- and fourth-largest communities $C_2$ and $C_4$).  We also use an intermediate connection strength within each member of the hierarchy, with  $J_{ee}^{in}\approx 0.83J_{ee}^{out}$ for connections within a cluster, $J_{ee}^{mix}=0.5J_{ee}^{in}\approx 0.42J_{ee}^{out}$ for neurons in distinct clusters but in the same hierarchy, and $J_{ee}^{out}$ the same as in Sec. \ref{Result1}.  This choice of $J_{ee}^{in}$ is lower than in Sec. \ref{Result1}, as the additional feedback from the hierarchy creates hyperactivity at $J_{ee}^{in}\approx 0.91J_{ee}^{out}$.   In order to produce a partially balanced hierarchical network with physically meaningful firing rates, we reweight the connections by the presynaptic community size as in Sec \ref{partialBalanceSec}:   a network with hierarchical community structure is generated using these parameters, and the connection strengths between clusters are rescaled to satisfy eq. \ref{actualBalance}.  Fig. \ref{fig:hierarchy}(A) shows there is no evidence of hyperactivity for the hierarchical network for these parameters, even though the network is not perfectly balanced ($\Wv'$ has non-negative eigenvalues).

To see the effect of partial balance on a network with a hierarchical community structure in response to a stimulus, we perform a simulation where 50\% of cluster 1 (in the first hierarchy) is directly stimulated over a time $0.5\mbox{s}\le t\le 2.5$s, followed by 50\% of cluster 2 (in the second hierarchy) being stimulated from $3\mbox{s}\le t\le 4$s.   In Fig \ref{fig:hierarchy}(B), we see that the stimulation of half of the neurons in both community 1 and 2 propagates within the community itself (consistent with Fig. \ref{fig:partial_balance_stim}).   Despite the significant differences in both duration of direct stimulation as well as the sizes of the underlying communities, activity in clusters 1 and 2 increases significantly (as shown in Fig. \ref{fig:hierarchy}(B)).  The firing rate, shown in the inset of Fig. \ref{fig:hierarchy}(B) (with blue circles for cluster 1 and red squares for cluster 2) is significantly greater for the hierarchical network than rate shown in Fig. \ref{fig:partial_balance_stim} due to the additional excitement feedback within the hierarchy.

The stimulation of clusters 1 and 2 not only excites activity in the intra-community neurons that do not receive direct stimulation (as was observed in Sec. \ref{partialBalanceSec}), but also propagates to the other clusters belonging to the same hierarchy (seen in Fig. \ref{fig:hierarchy}(B), indicated by the five- and six-pointed stars).  Stimulating a small portion of the neurons in a partially balanced hierarchy can thus produce synchronized firing in both member clusters in the hierarchy.  The activity in the un-stimulated cluster is comparable to that of the directly-stimulated cluster using our parameters), but further reduction of $J_{ee}^{mix}$ (decreasing the strength within the hierarchy) will reduce the downstream effects of stimulation of a member of the hierarchy.  Likewise, reducing $J_{ee}^{in}$ but keeping $J_{ee}^{mix}=0.5J_{ee}^{in}$ (reducing the degree of imbalance of the network $D(\Wv_p)$) will reduce the response of all communities to external stimulation (consistent with Fig. \ref{fig:densityplots}.  We also note that the stimulation effects persist even after the stimulation is turned off (visible in both Fig. \ref{fig:hierarchy}(B) and its inset), consistent with the observations in  \cite{LitwinKumar:2012go} (where stimulation persisted in homogeneous communities).  Reducing a smaller $J_{ee}^{in}$ or $J_{ee}^{mix}$ reduces this persistence time (data not shown), so the persistence of synchronized firing post-stimulation is dictated by the degree of imbalance.  Note that if $J_{ee}^{in}$ is increased that the stimulation time increases as well, leading to a long-lived hyperactive state.

\section{Conclusions}

In this article, we have looked at the effects of heterogeneous cluster sizes in a cortical network following the results of   \cite{LitwinKumar:2012go}. We found that bigger communities with the connection strengths similar to the homogeneous clusters can become hyperactive and suppress firing in the other communities altogether. This is not an expected behavior for real cortical networks. To remedy the effect of hyperactivity, one needs to enforce the balance condition on all communities of the network, which is done in Sec.~\ref{BalanceSec}. However, we found that a perfectly balanced network does not propagate stimulation as the balancing procedure gets rid of any community structure in the network. One must thus carefully introduce enough imbalance as to avoid hyperactivity but allowing synchronized firing and variability in the firing rate. In Sec.~\ref{Partial Balance}, we explain a procedure to quantify the imbalance that needs to be introduced in the network---thus producing a partially balanced network. We show that a partially balanced network can exhibit synchronized firing dynamics and can propagate stimulation within the community, thus restoring an active community structure. We also show that the same measure of imbalance $D(\Wv_p)$ produces similar firing dynamics in networks with different community structures. This is demonstrated in Fig.~\ref{fig:partial_balance_unstim}(B), where a firing dynamics similar to that of \cite{LitwinKumar:2012go} is obtained by ensuring both networks have similar values of $D(\Wv_p)$. 

In our article, we have focused on imposing partial balance on the network by redefining the connection strengths of the exciter neurons which form the communities. However, this is not the only method to create a balanced network. One could, in principle, modify the inhibitor connection strengths as well to get rid of the hyperactivity in the network. In the Supplementary Information, we show that it is possible to enforce the balance condition by changing the inhibitor connection strengths but it becomes exceedingly harder to do so as the number of communities in the network increases. The method used in our article is therefore more straightforward and easily scalable to networks with large number of communities.

We have also extended our results to networks with a hierarchical structure, where the method of partial balance ensures that stimulation propagates to communities within the same hierarchy. We have shown this for a simplified hierarchy as a proof-of-concept demonstration that our method can be extended to more complex networks with multi-layer top-to-bottom structure. Hierarchical structure in cortical networks is supported by anatomical and functional data, where different regions exhibit layered communication and asymmetric connectivity profiles, often linked to distinct timescales of activity and directionally organized signal flow   \cite{Markov:2013it,markov2014weighted,chaudhuri2015large}. The presence of hierarchy enables a network to maintain low spontaneous activity or quiescence under resting conditions while remaining sensitive to input. This is a direct consequence of local inhibitory-excitatory balance, which suppresses background firing but does not eliminate the potential for rapid activation upon stimulation. Our simulations demonstrate that quiescent dynamics and propagating activation are not incompatible states, but rather emergent features of the same underlying architecture and balance constraints. This separation of dynamical regimes enables cortical networks to remain energy efficient in baseline states while preserving responsiveness under task-specific activation. These results reinforce the functional relevance of both hierarchical organization and partial balance, and suggest a scalable control mechanism that can support large-scale and structured cortical dynamics.

\section*{Acknowledgements}
We acknowledge A. Litwin-Kumar and B. Doiron for providing the code to replicate their results in Ref.~\cite{LitwinKumar:2012go}. The code served as the template for our simulations. A.C. acknowledges the Department of Physics at University of Houston for the financial support during the completion of this work. G.M. acknowledges funding from the National Science Foundation, NSF-PHYS-2019745, as well as computational resources through NSF-CNS-1338099.

\putbib[Neural]
\end{bibunit}

\clearpage
\begin{bibunit}
\widetext
\begin{center}
\textbf{\large Supplementary Information}
\end{center}
\setcounter{section}{0}
\setcounter{equation}{0}
\setcounter{figure}{0}
\setcounter{table}{0}
\makeatletter
\renewcommand{\theequation}{S\arabic{equation}}
\renewcommand{\thefigure}{S\arabic{figure}}
\renewcommand{\thesection}{S.\Roman{section}}
\renewcommand{\bibnumfmt}[1]{[S#1]}
\renewcommand{\citenumfont}[1]{S#1}

	\section{Methodology}
	The foundation of the methodology used in our simulations is based closely on the parameters used in Ref.~\cite{LitwinKumar:2012go}. In this section, we briefly summarize the network model and parameters used in the simulations and how we modify the community sizes to create a network with a heterogeneous community structure. The codes used for the simulations in the main text can be found at this \href{https://github.com/abhijit975/Heterogenous-cortical-network}{{\color{blue}GitHub link}}.
	
	\subsection{Computational model}
	All simulated network consists of 4000 excitatory $(N_e)$ and 1000 inhibitory $(N_i)$ ($N = N_e + N_i = 5000$) leaky integrate and fire (LIF) neurons with their voltage (V) following the differential eq.(\ref{ODE_Voltage}). The numerical integration was done using the Euler method with time-step 0.1 ms and for a total of 4000 ms following Ref.~\cite{LitwinKumar:2012go}. The voltage of the $j$-th neuron at any given time $t$ is
	\begin{equation}
	\dot{V}_j (t)= \frac{1}{\tau_j}(\mu_j - V_j) + I_{j,syn}\;. \label{ODE_Voltage}
	\end{equation}
	Here, $\tau$ is the time constant of the membrane with $\tau_e = 15$ ms and $\tau_i = 10$ ms for an excitatory and inhibitory neuron, respectively. The threshold voltage when any individual neuron fires is set to $V_{th} = 1.0$ and after firing the resting voltage becomes $V_r = 0.0$ for a refractory period of 5 ms. $\mu_j$ is the bias voltage chosen from a uniform random distribution $\sim U(1.1,1.2)$ for an excitatory neuron and $\sim U(1.0,1.05)$ for inhibitory neurons. Even though the neuron is supra-threshold, the inhibitory synaptic currents ensure that the system is balanced \cite{LitwinKumar:2012go}. $I_{syn}$ is the synaptic input current to each neuron and is modeled by the equation,
	\begin{equation} 
	I_{j,syn}(t) = \sum_{k=1}^{N} J_{jk}\sum_{n} \alpha_k(t-t_{k,n}) \;,\label{syn_current}
	\end{equation}
	where $t_{k,n}$ is the $n$-th spike time of the $k$-th neuron, $J_{jk}$ is the strength of the connection from neuron $k$ to neuron $j$. $N$ is the total number of neurons in the network $(N= N_e+N_i)$. The function $\alpha(t)$ acts as a synaptic filter and is given by
	\begin{equation}
	\alpha(t) = \frac{1}{\tau_2 - \tau_1} (e^{-t/\tau_2} - e^{-t/\tau_1})\;,
	\end{equation}  
	with $\tau_2 = 3$ ms, $\tau_1 = 1$ ms for excitatory synapses and $\tau_2 = 2$ ms, $\tau_1 = 1$ ms for inhibitory synaptic connections \cite{LitwinKumar:2012go}.
	
	The probabilities of connections in the network with no community structures are $p_{ei} = p_{ie} = p_{ii} = 0.5$, where $p_{xy}$ denotes the connection probability between a neuron in population $x$ and a neuron in population $y$ with $x,y\in \{e,i\}$. Each excitatory neuron is connected to 800 other excitatory neurons. The degree of each neuron is the same and remains fixed in different simulations. For a network with no clusters, the network is in a balanced state (with excitement and inhibition approximately equal for each neuron). As expected, the network exhibits only random firing of the neurons---consistent with Fig. 1(d) of Ref.~\cite{LitwinKumar:2012go}. 
	
	To obtain more complex dynamics involving correlated firing, the excitatory neurons are divided into 50 communities of equal size, i.e., each community consisting of 80 excitatory neurons. The probability of connection and the connection strength inside the community are dictated by the network parameters $R_p$ and $R_J$ defined as
	\begin{equation}
	R_p = \frac{p_{ee}^{in}}{p_{ee}^{out}}\;\;,	\hskip 3em R_J = \frac{J_{ee}^{in}}{J_{ee}^{out}}\;. \label{ratios}
	\end{equation}
	In the simulation, to observe correlated behavior in individual communities, the parameters in Eq.~(\ref{ratios}) are chosen to be $R_p = 2.5$ and $R_J = 1.9$. The connection strengths are scaled as $\sim 1/\sqrt{K}$ where K = 800 is the degree of each excitatory neuron. In the units used, the connection strengths are $J_{ee}^{out} = 0.0236$, $J_{ie} = 0.0141$, $J_{ei} = -0.0453$, and $J_{ii} = -0.0566$. With these connection strengths and the parameter values $R_p = 2.5$ and $R_J = 1.9$, the results of the paper \cite{LitwinKumar:2012go} can be reproduced:  the simulated dynamics show variability in the firing rate and correlated dynamics within communities.  
		
	\begin{figure}[ht!]
		\centering
		\includegraphics[width=1.0\linewidth]{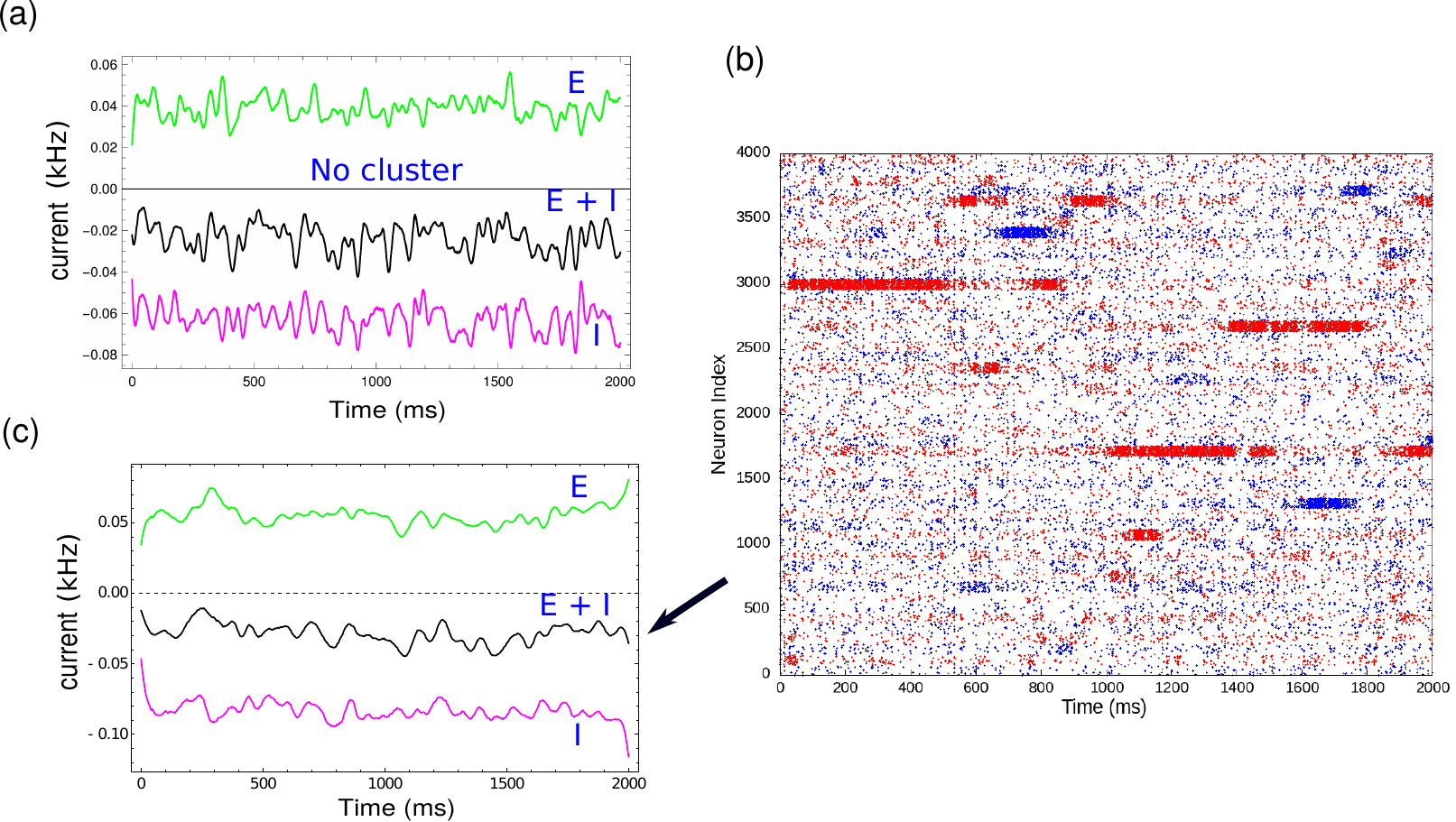}
		\caption{(a) Synaptic current in a network with no communities. The net current is in the inhibitory regime as it should be for a supra-threshold network. The solid line marked E (I) in green (magenta) denotes the excitatory (inhibitory) contributions to the total synaptic current E+I. The total synaptic current for the network with no communities can act as the baseline for determining whether a community in a network is hyperactive or not. (b) Spike raster plot shows correlated firing dynamics in a network with equal cluster sizes. An alternate coloring scheme differentiates between two neighboring communities. Each even-indexed community is shown in blue and odd indexed community is shown in red. (c) The corresponding synaptic current plot for the network in (b). Similar values of the total synaptic current of a randomly chosen community show that communities are not hyperactive even though they exhibit correlated firing dynamics.}
		\label{fig:homogeneousrastercurrent}
	\end{figure}
	
	\subsection{Checking balance condition}
	To check whether the balance condition in a network is maintained or not, we compare the total synaptic current (averaged over time) with that of a network with no clusters. As shown in Fig.\ref{fig:homogeneousrastercurrent}~(c), the same order of magnitude for the synaptic current with respect to the cluster-free network suggests that the network does not exhibit hyperactivity or hypersuppression. The synaptic current can be found out by plotting the quantity $I_{j,syn}$ for any neuron. There are two contributions to the synaptic current, received from the inhibitors and the exciters. These two contributions are also shown in Fig.~\ref{fig:homogeneousrastercurrent}, which demonstrate hyperactivity or hypersuppresion in the network.
	
	\section{Heterogeneous community sizes}
	\subsection{Gaussian heterogeneity}
\begin{figure}[htbp]
	\centering
	\includegraphics[width=1.0\linewidth]{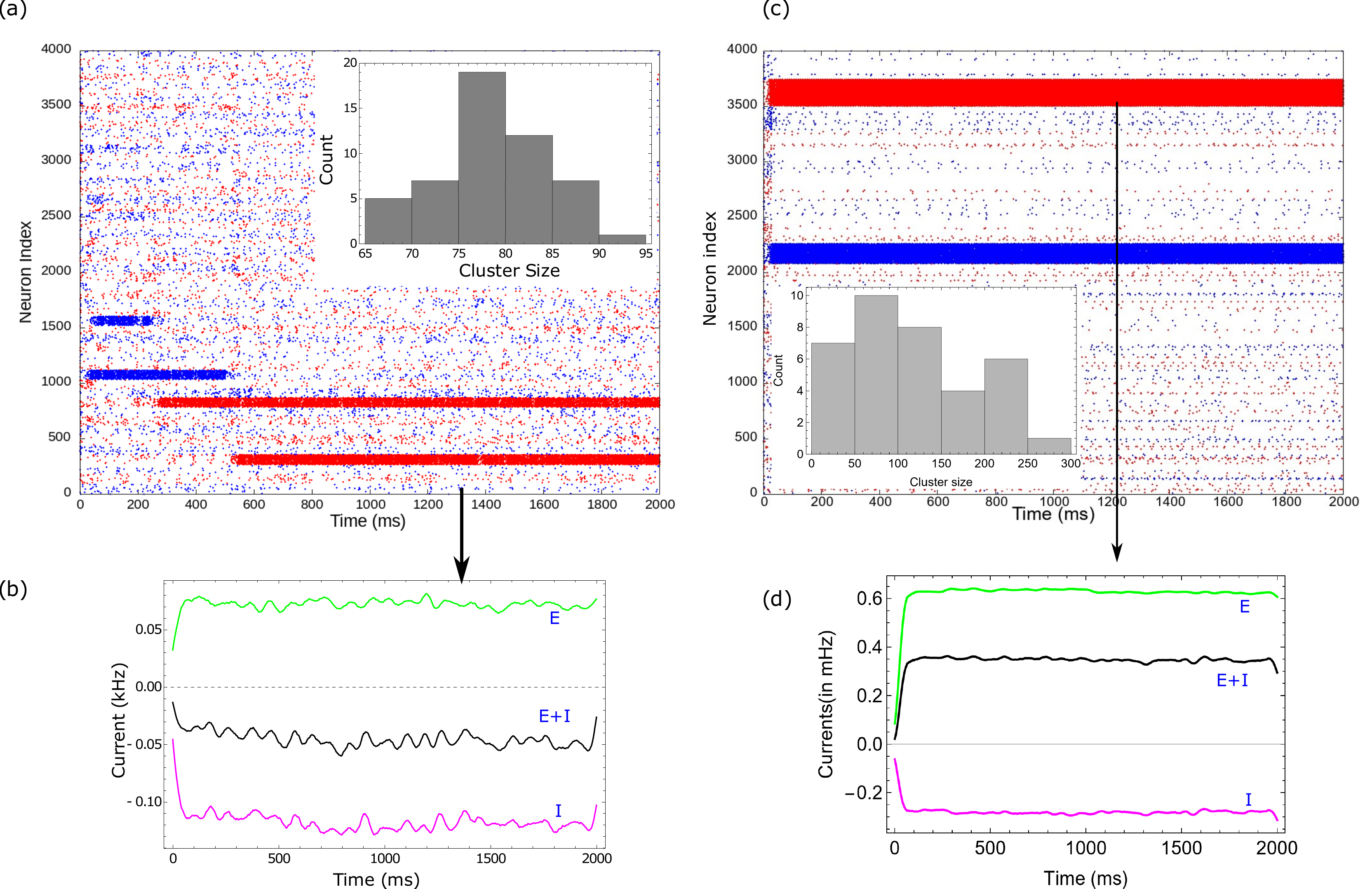}
	\caption{(a) Raster plot of a network with normally distributed cluster sizes demonstrates the robustness of balance condition with the introduction of mild heterogeneity. In the inset, the size distribution used for this particular raster plot is shown. (b) Corresponding synaptic current in an exciter belonging to the largest community is still comparable to the network with homogeneous community sizes. (c) Raster plot of a network with a normally distributed cluster size with large variation ($\sigma = 80$) shows hyperactivity and hypersuppression as the balance is broken due to increased heterogeneity in the community sizes.  (d) The corresponding synaptic current of a neuron in cluster 1 shows a large inhibitory current, which results in the suppression of firing in cluster 1.}
	\label{fig:gaussian}
	\end{figure}

	In the previous section, the cluster sizes in the exciter population were exactly the same---leading to a network with homogeneous clusters. However, homogeneity in community sizes is not guaranteed for real cortical networks.  In many biologically relevant examples\cite{deReus:2014cz,Klinshov:2014ce}, heterogeneity in the size of clusters of neurons has been observed.  To model such heterogeneity,  we would like to see what happens when the community sizes in the network vary. One way to incorporate heterogeneity is to make the cluster sizes follow a Gaussian distribution. The mean community size is maintained at $\langle c \rangle = 80$ (same as the network with homogeneous communities) but with a standard deviation of $\sigma$ which can be large $(\sigma/\langle c \rangle>1)$ or small $(\sigma/\langle c \rangle<1)$.
	For $\sigma = 6$, the typical community sizes range between $60-100$, and the effect of heterogeneity is mild as seen in Fig.~\ref{fig:gaussian}(a). The stochastic correlation is still observed in the small communities with increased activity in the larger communities. This is consistent with the results of Ref.~\cite{LitwinKumar:2012go}. Note that the total synaptic current in Fig.~\ref{fig:gaussian}(b) for the network with normally distributed community sizes with small variance is not large compared to the homogeneous network in Fig.~\ref{fig:homogeneousrastercurrent}(c), indicating small heterogeneity has little effect on the firing dynamics.
	
	For large variance (e.g., $\sigma =100$) of community sizes, we cut off the Gaussian distribution below $c=0$, which produces firing activity consistent with Sec.~2 of the main text---large communities are more likely to become hyperactive, suppressing the firing in all other communities as shown in Fig.~\ref{fig:gaussian}(c) and (d). This is because, for large variance, we can get relatively large communities, which is comparable to the large community sizes in the network used for the simulations in the main text.
	
	\subsection{Exponential distribution \label{expdist}}
	The problem with Gaussian distribution is that for larger variance, there is an increased chance of getting negative community sizes. Getting rid of negative community sizes modifies the probability distribution from which sampling is done. So, we choose a distribution that always yields positive numbers for community sizes, i.e. the exponential distribution. We create a network of $N_e = 4000$ neurons with the community sizes following the exponential distribution $P(c)\sim e^{-c/\lambda}$ with  $\lambda = 80$ fixing the mean $\langle c \rangle \approx 80$ (the same as in Ref. \cite{LitwinKumar:2012go}) and standard deviation $ \sqrt{\langle c^2\rangle-\langle c\rangle^2}\approx 80$ (a nonzero variance). To generate a network, we randomly choose the size of each community $c_i$ with probability $e^{-c_i/\lambda}(e^{1/\lambda}-1)$.  After the $K^{th}$ community is added, we compute the number $N(K)=\sum_{k=1}^Kc_k$ and determine if more communities are to be added.  If $N(K)<4000$, we add a new community (setting $K\to K+1$), if $N(K)=4000$, we have added the precise number of desired excitatory neurons and the community structure is accepted, and if $N(K)>4000$ we set $K=0$ and begin again from scratch.  This procedure produces $C \approx 50$ clusters comparable to the homogeneous cluster size cases. With this distribution, the largest cluster contains about $\sim 500$ neurons, and the smallest cluster can contain as low as 2 neurons, with a few individual neurons that do not belong to any cluster (or equivalently where $c=1$). This network is used in the main text where the resulting hyperactivity has been shown explicitly in Fig.~2 of the main text. 

	\subsection{Power-law distribution}
	The power-law distribution ($P(c)\sim c^{-a}$) is another significant distribution since scale-free networks have been observed in many real-world networks \cite{albert2005scale,barabasi2000scale,eguiluz2005scale,ravasz2002hierarchical,stam2004scale}. However, a power-law distribution may not have a well-defined mean (for $a<1$) or variance (for $a<2$). Scale-free networks with the desired mean will generally result in a network with a large number of very small communities and a few very large communities.  We found empirically that a power-law distribution with exponent $a = 1.5$ produces on average $\langle C\rangle \approx 50$ clusters with the constraint $N_e = 4000$, and shows the same behavior (hyperactivity) as discussed in the main text.  However, noting that an exponent $a=1.5$ has a diverging variance for $N_e \rightarrow \infty$, we choose not to focus on this distribution in the main text.  For individual realizations of a scale-free distribution of community sizes, we expect that the procedure to restore balance described in this paper will be applicable.

	\section{Synchronization}
\begin{figure}[!htbp]
		\centering
		\includegraphics[width=0.9\linewidth]{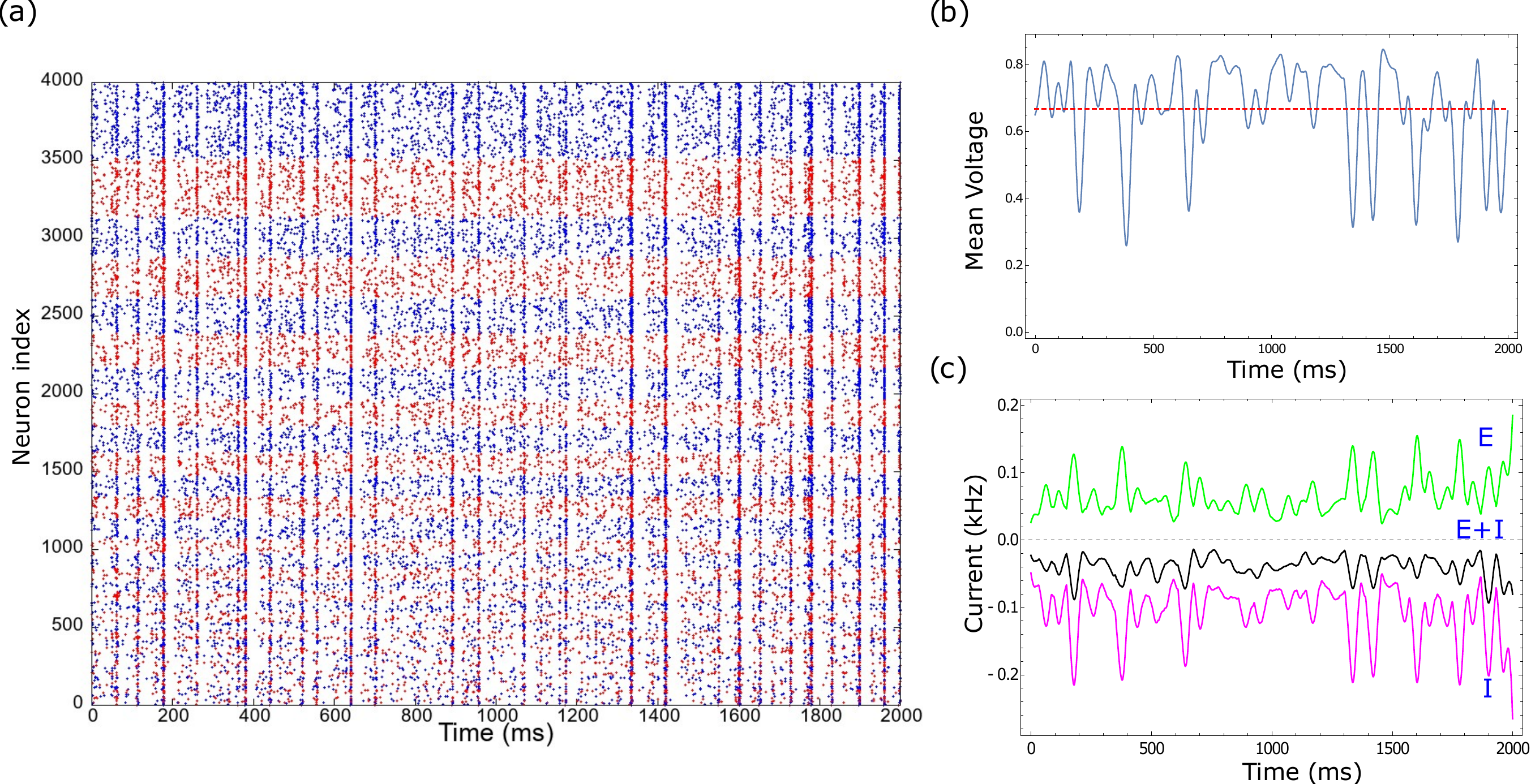}
		\caption{(a) Raster plot of the network after the balancing procedure has been applied to the whole network with largely heterogeneous cluster size following an exponential distribution. The spikes show synchronized behavior throughout the network, irrespective of the clustering. (b) The corresponding mean voltage (averaged over all exciter neurons) plot with time shows the mean voltage of the exciters fluctuating around the temporal average of the mean voltage. Each fluctuation indicates synchronized firing across the network and acts as a measure of the synchronized behavior. The red dashed line shows the temporal average of the mean voltage. (c) Synaptic current in one representative neuron in cluster 1. The synaptic current shows the fluctuations corresponding to the globally synchronized firing in the network.}
		\label{fig:syncexp}
	\end{figure}

	After a network with exponentially distributed cluster sizes has been created, we simulate the network by performing the balancing procedure upon the whole network as prescribed in Sec.~3 of the main text, with $C^*=C$ (with the strength of all excitatory connections scaled by the size of the communities) and observed a globally synchronized dynamics in the network as shown in Fig.~\ref{fig:syncexp}(a). To quantify the degree of synchronization, we used an order parameter 
	\begin{equation}
	m = \log\bigg[Var\bigg (\frac{1}{N_e}\sum_{i=1}^{N_e}V_i \bigg )\bigg] = \log\big[Var\big(\langle V \rangle_t\big)\big]\;, \label{order_parameter}
	\end{equation}
	which has been used previously in other studies of synchronization in neural networks \cite{kim2015thermodynamic}. Here, $\langle V \rangle_t$ is the mean voltage of all exciter neurons at a particular time $t$ and the variance is over time.	For a network with no synchronized dynamics and irregular random firing, the average voltage of the neurons remains almost at the same level, with very small fluctuations from the mean value. For synchronized dynamics, when a large fraction of the neurons fire together, there is a surge in the average voltage followed by a dip. So, the average voltage of the neurons shows an oscillatory behavior around the mean value with large fluctuations (see Fig.~\ref{fig:syncexp}(b) and (c)). The larger the fluctuations, the more number of neurons are firing together, indicating a greater synchronization.  Using the order parameter in Eq.~(\ref{order_parameter}), we can quantify the presence of synchronization in the network. For a network with homogeneous cluster sizes, we obtain $m = -6.22\pm 0.01$, whereas, for the exponentially distributed communities, we get $m= -3.3 \pm 0.01$.  The synchronization is thus significantly larger for a network with exponentially distributed community sizes compared to a homogeneous network, after the balancing procedure is performed on the whole network. 
	
	\section{Disappearance of Global Synchronization}\label{sec:disapp_sync}
	Even though the procedure described in Sec.~3 of the main text creates a balanced weight matrix for a largely heterogeneous network with the total connection strength remaining the same, it attributes a large connection strength to the smaller communities. For example, neurons belonging to a community of 2 neurons will have 250 times the connection strength of the neurons belonging to a community of size 500. This results in the small communities dominating the dynamics of the whole network. If the factor $\varphi(C)$ is sufficiently large, the smaller communities can trigger the inhibitors to fire whenever they are firing, which, in turn, suppresses the other exciters in the network resulting in the simultaneous inactivity in the whole network. These stripes of inactivity, followed by coherent firing, create the apparent synchronized behavior in the heterogeneous networks. 
	
	Physically, we expect a small number of neurons should not affect the global dynamics of the network. Such a neural network would be highly sensitive to the dynamics of a few individual neurons and weakly sensitive to clusters of hundreds of neurons.  We expect that meaningful neural networks modeling the brain should be robust to the behavior of a few unbalanced neurons. So, we introduce a method that ensures that the large communities are balanced using the procedure described in Sec.~3 of the main text, but the strengths of the smaller communities are kept unchanged. The division between `large' and `small' is unclear, and, in this section, we develop a method to determine the cutoff between the two groups. The goal is to identify the largest subset of neurons whose effect is negligible, in the sense that its effect on the balanced state of the other neurons is bounded in the limit of $t\to\infty$.
	\subsection{Threshold functions}
	A commonly used approximation\cite{vreeswijk1998chaotic,dayan2001theoretical,ermentrout2010mathematical} for the dynamics of the firing rate of a collection of neurons is $\tau \dot\rv=-\rv+f_s(\Wv^\prime\rv+b)$ where $b$ is a bias term, $\Wv^\prime$ the matrix of connection strengths between neurons (node to node) and $\rv$ is a vector of firing rates of each neuron.  This can be coarse-grained to the connections between communities \cite{Pyle:2016bvb} (as was done in Sec.~3 of the main text).  We model the response using a sigmoid function, with $f_s(x)=f_0(1+\tanh(x/s))/2$ where $s$ is the width of the transition \cite{ginzburg1994theory} and $f_0$ is a constant. $f_s(x)$ has vanishing contribution for $x\ll 0$ and saturates when $x\gg 0$.  For large $x$, $f_s'(x)\to 0$ exponentially fast. 
	\subsection{Leading order effects}
	Prior to the addition of the perturbation, we assume we have a balanced network of $N_b$ nodes with firing rates $\bar\rv(t)$ satisfying $\tau\dot{\overline{\rv}}=-\overline{\rv}+f_s(\overline{\Wv}\overline{\rv}+\overline{\bv})$.  These nodes are assumed to be divided into $C^*$ communities with any distribution of size.  After the perturbation is added, we divide the neural firing rate into a vector of length $N$ of the balanced neurons, with firing rate $\bar\rv+\xv$, and a vector of length $N_y$ of the (possibly unbalanced) perturbation, $\yv$.  We rewrite the connectivity matrix as
	\begin{eqnarray}
	\Wv=\left(\begin{array}{cc}\overline\Wv & \Av \\\Bv & \Cv\end{array}\right)
	\end{eqnarray}
	for the submatrices $\Av$ (an $N_b\times N_y$ matrix), $\Bv$ ($N_y\times N_b$), and $\Cv$ ($N_y\times N_y$) representing the connection strengths between the two divisions.  Assuming $x_i$ is small (meaning the rates are weakly perturbed by the new nodes) and $\Av\yv$ is small in comparison to $\overline{\Wv}\overline{\rv}$ (which assumes  the dynamics of the balanced network is dominated by the firing within it, not the firing of the perturbation), we can write approximately
	\begin{eqnarray}
	\tau \dot x_i&\approx&-x_i+ f_s'\bigg[(\overline{\Wv}\overline{\rv})_i+b_i\bigg]\bigg(\overline{\Wv}\xv+\Av\yv\bigg)_i\\
	&\equiv&-x_i+\phi^x_i(t)\bigg(\overline{\Wv}\xv+\Av\yv\bigg)_i\label{firsteq}\\
	\tau \dot y_i&\approx&-y_i+f_s'\bigg[(\Bv\overline{\rv})_i+b_i\bigg]\bigg(\Bv\xv+\Cv\yv\bigg)_i+f_s\bigg[({\Bv}\overline{\rv})_i+b_i\bigg]\\
	&\equiv&-y_i+g_i(t)+\phi^y_i(t)\bigg({\Bv}\xv+\Cv\yv\bigg)_i\label{secondeq}
	\end{eqnarray}
	where we have defined the auxiliary functions
	\begin{gather}
	g_i(t)=f_s\left[(\Bv\overline\rv(t))_i+b_i\right]\;,\\ \phi^x_i(t)=f_s'\left[(\overline{\Wv}\overline\rv(t))_i+b_i\right]\;,\\
	\phi_i^y(t)=f_s'\left[(\Bv\overline\rv(t))_i+b_i\right]
	\end{gather}
	for convenience, depending solely on the unperturbed value of $\overline{\rv}(t)$.  While a limit cycle is possible within the unperturbed network, the analysis is significantly more complicated and it is convenient to assume that in the limit of $t\to\infty$ the unperturbed network reaches a steady state with $\overline\rv(t)\to\overline\rv_\infty$.  
	\subsection{Feedback in perturbations composed of exciters}
	While $ (\overline{\Wv}\overline{\rv})_i$ is small for balanced networks at steady state (that is, the input from exciters is balanced by the input of the inhibitors on average), there is no similar constraint on $\Bv\overline\rv_\infty$ (the effect of the original network on the perturbation).  If we assume the perturbation is composed entirely of exciters (as is the case for our original problem of the division between large and small communities of exciters), the greatest effect the perturbation can have is in the case where $(\Bv\overline\rv+\bv)_i$ is large for all $i$ and $f_s'(\Bv\overline\rv+\bv)$ reaches its saturating value. Using the sigmoidal function, $f_s'(\Bv\overline\rv+\bv)\approx 0$ and $f_s(\Bv\overline\rv+\bv)\approx f_0$ in this limit and we find $y_i\approx f_0$ for all $i$ at steady state.  Defining $\Phiv_{ij}^x=\delta_{ij}\phi_i^x(t) = \delta_{ij}f_s'\left((\overline \Wv\rv_\infty)_i+b_i\right) = \delta_{ij} f_s'\left(\overline{I}_{syn}\right)$, the effect on the original network is
	\begin{eqnarray}
	\xv_s\approx\Phiv^x(\onev-\Phiv^x\Wv)^{-1}\Av\fv_0\label{sigmoideq}
	\end{eqnarray}
	with $(\fv_0)_i=f_0$.  Convergence of $\xv$ is guaranteed so long as $\Phiv^x\Wv$ has all eigenvalues less than one, which sets a minimal condition on maintaining balance for the sigmoid function.
	\subsection{Procedure for generating partially balanced networks}
	The effect of the perturbation on the original network, quantified by $\xv$ can be determined exactly to first order but depends on the (unknown apriori) values of $\rv_\infty$.  To balance the larger communities in the heterogeneous network while treating smaller communities as a perturbation, we implement the following procedure:
	\begin{enumerate}
		\item We initially take the two largest communities as our primary network and all others as a perturbation.   In case of a tie in size, we randomly choose two communities.  The weights in the primary network are scaled so that the network is balanced as described in Sec.3 of the main text, and the weights in the perturbation are not altered.  We compute the normalizing factor in Eq.~(8) of the main text using 
		\begin{equation}\label{factor_partial}
		\varphi(C^*) = \frac{[R_p R_J(N_0-1)+(N_e-N_0)]N_e - \sum_{k=1}^{C'}[R_p R_J (N_k-1)+(N-N_k)]N_k}{\sum_{k=C'+1}^{C} R_p R_J^{\prime}(N_k-1)+(N-N_k)}
		\end{equation}
		where $C'=C-C^*$ is the number of communities left unbalanced. We also choose a tolerance $\delta$ as well as the width $s=1.0$ and saturating values $f_0=1$ for the threshold functions.
		\item We compute $\rv_\infty$ for the primary network, and determine $\xv_s$ in Eq.~(\ref{sigmoideq}).
		\item From this, we compute $x^2=\xv_s^T\xv_s$.  If $x\le \delta$, we halt and use these weights.  If $x>\eps$, we add the next-largest community to our primary network, reweigh the edges in the primary network, and return to step 2.
	\end{enumerate}
	The end result of this is a balanced network of large communities, connected to unbalanced small communities that do not disrupt the balance of the primary network.  
	\subsection{Result} 
\begin{figure}
		\centering
		\includegraphics[width=1.0\linewidth]{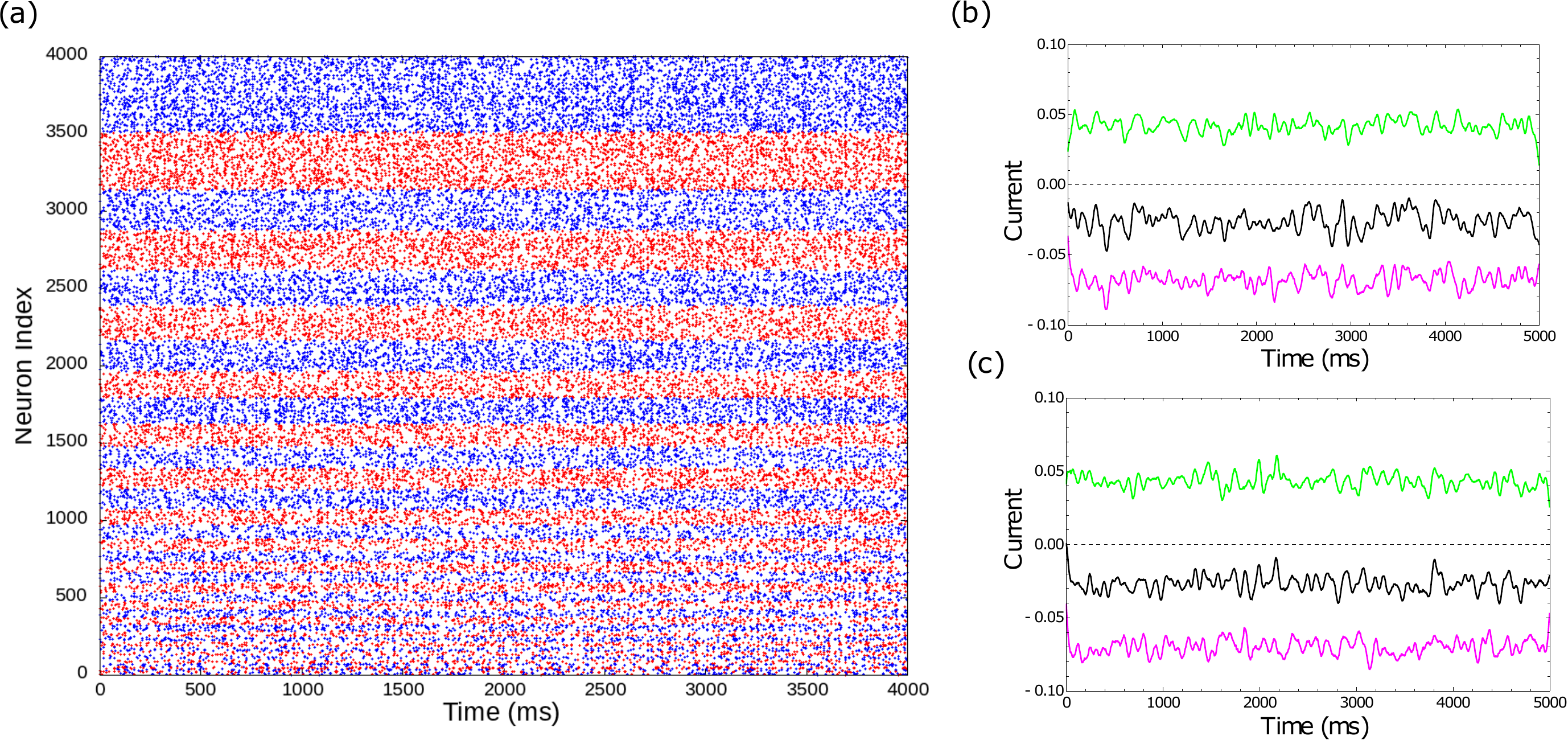}
		\caption{(a) Raster plot of the rebalanced network following procedure described in Sec.~\ref{sec:disapp_sync}. The plot does not indicate any hyperactive or suppressed community. (b) and (c) There is no excessive excitatory or inhibitory net synaptic current in clusters 1 (top) or 2 (bottom), respectively.}
		\label{fig:nosynccolor}
	\end{figure}

	Applying this procedure to a network of 4000 excitatory and 1000 inhibitory neurons with the excitatory population forming clusters with exponentially distributed size yields a network that is still balanced and does not exhibit synchronization (Fig.~\ref{fig:nosynccolor}). In this procedure, the values of the used parameters are $f_0 = 1$, $s = 1.0$, and the tolerance $\delta = 0.10$. Using these values, the largest size of the community that was dropped from the primary network is 25.  In the main paper, we use this cutoff of $c_{min}=25$ for all exponential networks that are generated and rebalanced.  The absence of globally synchronized behavior indicates that this procedure removes the enormous connection strengths of small communities, as expected.
	
	\section{Discussion on Inhibitors}
	\subsection{Rebalancing a network by adjusting inhibitory strengths}
	The solution of re-scaling the connection strengths between exciters is not the only solution to recover a perfectly balanced network (as described in Sec.~3 of the main text.  Reweighing the exciter strengths effectively removed the community structure of the weighted network (by imposing a between-community strength greater than the within-community strengths).  An alternate approach to re-balancing the network would preserve the exciter strengths and community structure but re-scale the inhibitor strengths to prevent overactivity in any exciter community. For the simple case of two exciter communities, we consider a weight matrix
	\begin{eqnarray}
	\Wv=\left(\begin{array}{ccc}N_1 a & N_2 b & -M c_1 \\N_1 b & N_2 a & -M c_2 \\N_1 d & N_2 d & -M e\end{array}\right)\label{exampleweights}
	\end{eqnarray}
	where we impose the constraints $a,b,c_i,d,e>0$.  A balanced matrix will have all negative eigenvalues, which can be imposed by rescaling the weights of each excitatory link using the size of the community, as discussed in the main text.  In this section of the SI, we instead consider varying $c_i,d,e$ while having the additional constraint $a<b$ (as in the text).  In the main text, $a$ and $b$ were modified to satisfy balance, but in this section, we will hold $a$ and $b$ fixed and vary $c_i$ and $e$ in order to balance the matrix.  This amounts to choosing $c_i$ and $e$ so that the matrix has all negative eigenvalues.  In order to do so, we will select positive real numbers $\lambda_1,\lambda_2,\lambda_3>0$, and determine the values of $c_i$ and $e$ that fix the eigenvalues of the weight matrix at $\{-\lambda_i\}$.  This is accomplished by requiring $p(\lambda)=(\lambda+\lambda_1)(\lambda+\lambda_2)(\lambda+\lambda_3)$ with $p(\lambda)=|\Wv-\lambda\Ov|$ the characteristic polynomial for the matrix.   Equating the coefficients allows us to find a simple condition, $e=(a+\lambda_1+\lambda_2+\lambda_3)$.  This means that $e>0$ for any choice of our parameters (since $a$ and $\lambda_i$ are already constrained to be positive).  This somewhat simple expression for $e$ must be combined with more complicated expressions for $c_i$'s:
	\begin{eqnarray}
	c_1=\frac{-g_0-g_1\delta N+ g_2\delta N^2+g_3\delta N^3}{8(a-b)dMN_1\delta N}\qquad c_2=\frac{-g_0-g_1\delta N+g_2\delta N^2-g_3\delta N^3}{8(a-b)dMN_2\delta N}\label{ceq}
	\end{eqnarray}
	with $\delta N=N_1-N_2$, $\delta w=b-a>0$, $D=\prod_i\lambda_i>0$, $T=\sum_i\lambda_i>0$, $C=\lambda_1\lambda_2+\lambda_2\lambda_3+\lambda_1\lambda_3>0$, 	 and where 
	\begin{eqnarray}
	g_0&=&\prod_i[\delta w N_e-2\lambda_i]\\
	g_1&=&\delta w[(3 a^2 + b^2)N_e^2 +4 C + 4 a N_e T]\\
	g_2&=&(a^2-b^2)[(3a-b)N_e+2T]\\
	g_3&=&(a+b)\delta w^2
	\end{eqnarray}
To ensure a balanced network has been created, it must be that $c_i>0$, and acceptable values of $\lambda_i$ satisfying this condition depends on the choice of $a$ and $b$, $N_i$ and $M$.  We note that the inhibitor-to-exciter strength $d$ enters only in the denominators in Eq.~(\ref{ceq}) (not in the factors of $\{g_i\}$).  Since $d>0$ this implies if a balanced matrix exists for any $d$, given values of $a$, $b$, and $\delta N$, the network will remain balanced for all other values of $d$ regardless of any change to $c_i$ or $e$.  Thus, adjusting the weight $d$ is unnecessary to ensure a balanced network for $a>b$, and we need to simply focus on varying the inputs to the inhibitory neurons.  

We use Mathematica's Reduce to determine the parameters that allow a simultaneous condition of all-negative eigenvalues in the weight matrix as well as all negative values for the last column of the weight matrix (due to the exciters). We find that there exist some inhibitory weights for which the matrix can be balanced holding $a$, $b$, $d$, $N_i$ fixed. One example is a family of solutions that simultaneously satisfy
\begin{eqnarray}
\frac{1}{a-b}\bigg(a+\sqrt{\frac{b}{a+b}\bigg[4a^2-3ab+b^2\bigg]\ }\bigg)&\le& \frac{\delta N}{N_e}\le 1\\
\frac{1}{2}\bigg[b-a+(b+a)\frac{\delta N}{N_e}+\sqrt{2b(a+b)\frac{\delta N}{N_e}\bigg(1+\frac{\delta N}{N_e}\bigg)}\bigg]&\ge& \lambda_1\ge \lambda_2\ge \lambda_3>0 \label{foundsoln}
\end{eqnarray}
Any choice of $\lambda_i$ satisfying these constraints will produce a balanced matrix so long as $a<b$.  This is one of the simpler and more easily expressed constraints on the values of $\{\lambda_i\}$, and is not exhaustive.   Note that the complexity of the specific solution derived in Eq.~(\ref{foundsoln}) indicates the difficulty of constructing a balanced network by adjusting inhibitory strengths, even in the relatively simple case of only two heterogeneous communities.

\subsection{The impossibility of rebalancing through inhibitory-to-excitatory connections}

\begin{figure}[htbp]
\includegraphics[width=1.0\linewidth]{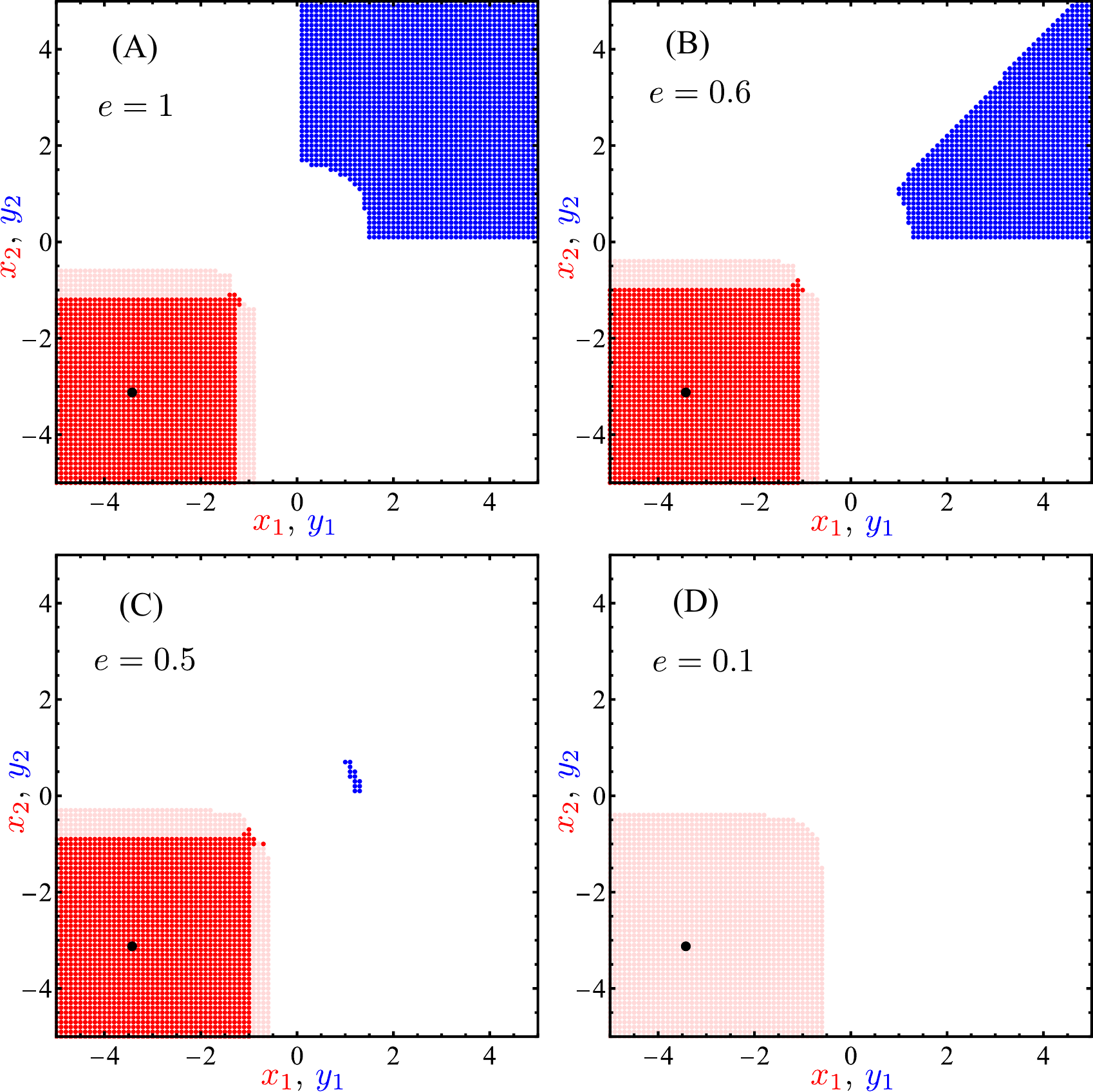}
\caption{Parameters producing balanced networks either by varying $x_i$ (exciter-to-exciter strengths) and $y_i=1$ (inhibitor-to-inhibitor strengths), shown in the red points, or by varying $y_i$ and $x_i=1$, shown in the blue points.  Each panel has the same values for $N_i$, $M$, $a,b,c$, and $d$, but have varying inhibitor-to-inhibitor strength $e$, with $e=1$ for (A), 0.6 for (B), 0.5 for (C), and 0.1 for (D).    Dark red and dark blue points indicate negative purely real eigenvalues and light red and light blue indicate complex eigenvalues with negative real values.   The black point indicates the solution of $x_1=1/N_1$ and $x_2=1/N_2$ as described in the main text.  In panel D the blue points have vanished while the red points persist, indicating rebalancing of the excitatory links ensures balance over a wider parameter space.  }
\label{fig:exciteVsInhibit}
\end{figure}

In the main text, we showed that choosing the excitatory interaction strengths inversely proportional to community size would produce a balanced network for heterogeneous connectivity (with the inhibitory strengths independent of community size).  In eq. \ref{foundsoln}, we found that we could have equally well held the excitatory connections constant and adjusted the inhibitory strengths to ensure balance (at least for a network with $C=2$ communities with heterogeneous size).  One might naturally wonder whether there is an advantage to focusing on adjusting the excitatory strengths (as we have done in the main text) over adjusting the inhibitory links in order to ensure balance.  We have argued that the simple scaling rule of $J'_{jk}\propto J_{jk}/N_k$ is sufficiently simple to be easily and usefully applied to highly heterogeneous networks, while the solutions in Eq.~(\ref{foundsoln}) are arguably more complicated. 

To better understand the utility of adjusting exciter-to-exciter strengths in comparison to inhibitor-to-exciter strengths, we chose a modified rebalance of the matrix in eq. \ref{exampleweights} by writing
\begin{eqnarray}
\Wv'_{mod}=\left(\begin{array}{ccc}N_1 a x_1 & N_2 b x_2 & -M c y_1 \\N_1 b x_1 & N_2 a x_2 & -M c y_2 \\N_1 d & N_2 d & -M e\end{array}\right)
\end{eqnarray}
This rescaling holds $d$ and $e$ fixed, and permits variation of the exciter-to-exciter (via $x_i$) or inhibitor-to-inhibitor (via $y_i$) strengths.  Note that this differs from the methodology in the main text (where $d$ was rebalanced as well), a choice made to restrict the variations in the parameter space to two dimensions.    In SI Fig. \ref{fig:exciteVsInhibit}, we show the values of $x_i$ or $y_i$ that produce a balanced network with the choices $M=1000=\frac{3}{8}N_1=\frac{3}{4}N_2$, $a=\frac{1}{2}b=c=d=0.1$, and varying $e$.  Note that this has imposed the constraint that $a<b$, but that the network is not balanced when $x_i=y_i=1$.  Red points indicate values of $x_i$ that produce a balanced network (all eigenvalues of $\Wv'_{mod}$ having negative real part) when $y_i=1$, and blue points indicate values of $y_i$ that produce a balanced network when $x_i=1$.  Fig. \ref{fig:exciteVsInhibit} clearly shows that it is {\em{always}} possible to adjust the excitatory strengths to ensure balance (for these particular choices of the base parameters $N_i,$ $M$, and $a,b,c,d,e$), but that the solutions for $y_i$ satisfying balance only {\em{sometimes}} exist.   For large $e$ (the inhibitory-to-inhibitory strength), a balanced solution is possible for a wide range of $y_i>0$, but for smaller values of $e$ (near $e=0.5=5a$) there are severe restrictions on the choices of $y_i$ that will produce a balanced network.  For sufficiently small $e$, no such solution exists.  Varying the particular base parameters $N_i$, $M$, and $a-e$ produces the same qualitative behavior.  Producing a balanced network by adjusting the inhibitor-to-exciter strengths is thus far more sensitive to the particular parameters in the model, justifying the focus on exciter-to-exciter links described in the main text.  
\putbib[Neural]
\end{bibunit}

\end{document}